\documentclass[%
 reprint,
 amsmath,amssymb,
 aps,
 prapplied
]{revtex4-2}

\usepackage{xcolor}
\usepackage{amsmath}
\usepackage{graphicx}% Include figure files
\usepackage{dcolumn}% Align table columns on decimal point
\usepackage{bm}% bold math
\usepackage{fancyhdr}

\begin{document}

\preprint{APS/123-QED}

\title{One-transmitter Multiple-receiver Wireless Power Transfer System Using an Exceptional Point of Degeneracy}% Force line breaks with \\
%\thanks{}%

\author{Fatemeh Mohseni}
% \email{fatemem@uci.edu}
\affiliation{Department of Electrical Engineering and Computer Science, University of California, Irvine, California 92697, USA}%Lines break automatically or can be forced with \\

\author{Amin Hakimi}
% \email{amin.hakimi@uci.edu}
\affiliation{Department of Electrical Engineering and Computer Science, University of California, Irvine, California 92697, USA}

\author{Alireza Nikzamir}
% \email{anikzami@uci.edu}
\affiliation{Department of Electrical Engineering and Computer Science, University of California, Irvine, California 92697, USA}

\author{Hung Cao}
% \email{hungcao@uci.edu}
\affiliation{Department of Electrical Engineering and Computer Science, University of California, Irvine, California 92697, USA}

\author{Filippo Capolino}
\email{f.capolino@uci.edu}
\affiliation{Department of Electrical Engineering and Computer Science, University of California, Irvine, California 92697, USA}

\begin{abstract}
Robust transfer efficiency against the various operating conditions in a wireless power transfer system remains a fundamentally important challenge. This challenge becomes even more critical when transferring power to groups of inductively coupled receivers. We propose a method for efficient wireless power transfer to multiple receivers exploiting the concept of exceptional points of degeneracy (EPD) in a self-oscillating system. In previous studies based on PT symmetry, a receiver’s operation has been divided into two strong and weak coupling regimes, and the power transfer efficiency is constant in the strong coupling regime when varying the coupling factor. Here, the concept of strong and weak coupling and constant power efficiency is extended to a system of multiple receivers that do not follow PT symmetry. The transmitter has saturable nonlinear gain and the EPD is analyzed when the system oscillates after reaching saturation.  We show that the important feature to have a roughly constant power efficiency, independently of the positions of the receivers, is the existence of an EPD that separates the weak and strong regimes. Our proposed method demonstrates a system with less sensitivity to the coupling change than a conventional system without EPD when the receivers and their couplings to the transmitter are not necessarily identical.
% Keywords-Wireless Power Transfer; Exceptional Points of Degeneracy; Inductive Coupling.

\end{abstract}

\maketitle

\section{Introduction}
Wireless power transfer (WPT) is a technology that enables the transmission of electrical energy without physical connectors, relying on electromagnetic fields to transfer power between a transmitter and receiver. It has paved the way for recent advances in portable electronic devices in a variety of applications, from eliminating the dominance of the battery on biomedical devices to wirelessly powered laptops needless of charging wires \cite{sample2010analysis, song2021wireless, hui2023wireless}. Methods can be categorized into two main classes: radiative and non-radiative WPT. Like far-field or RF broadcast techniques, radiative WPT is based on propagating electromagnetic waves carrying energy. Although they are suitable for transferring data, there are certain disadvantages for transmitting power, e.g., in far-field approaches, the efficiency depends on the transmitter's directivity. Although in RF broadcasting methods, this is solved by omnidirectional propagation, the transferred power drops rapidly by increasing distance due to the $1/r^{2}$ dependency. Therefore, non-radiative methods like inductive coupling based on resonant LC tanks are becoming more widespread \cite{
kurs2007wireless,liu2023smart}. Nowadays, inductive WPT can be found in a variety of devices, such as biomedical implants \cite{hajiaghajani2019patterned}, electric automobiles \cite{sakhdari2020robust}, household appliances \cite{shu2018wireless}, and laboratory research\cite{mohseni2021design}.

\begin{figure}
\centering
\includegraphics[width=1\columnwidth]{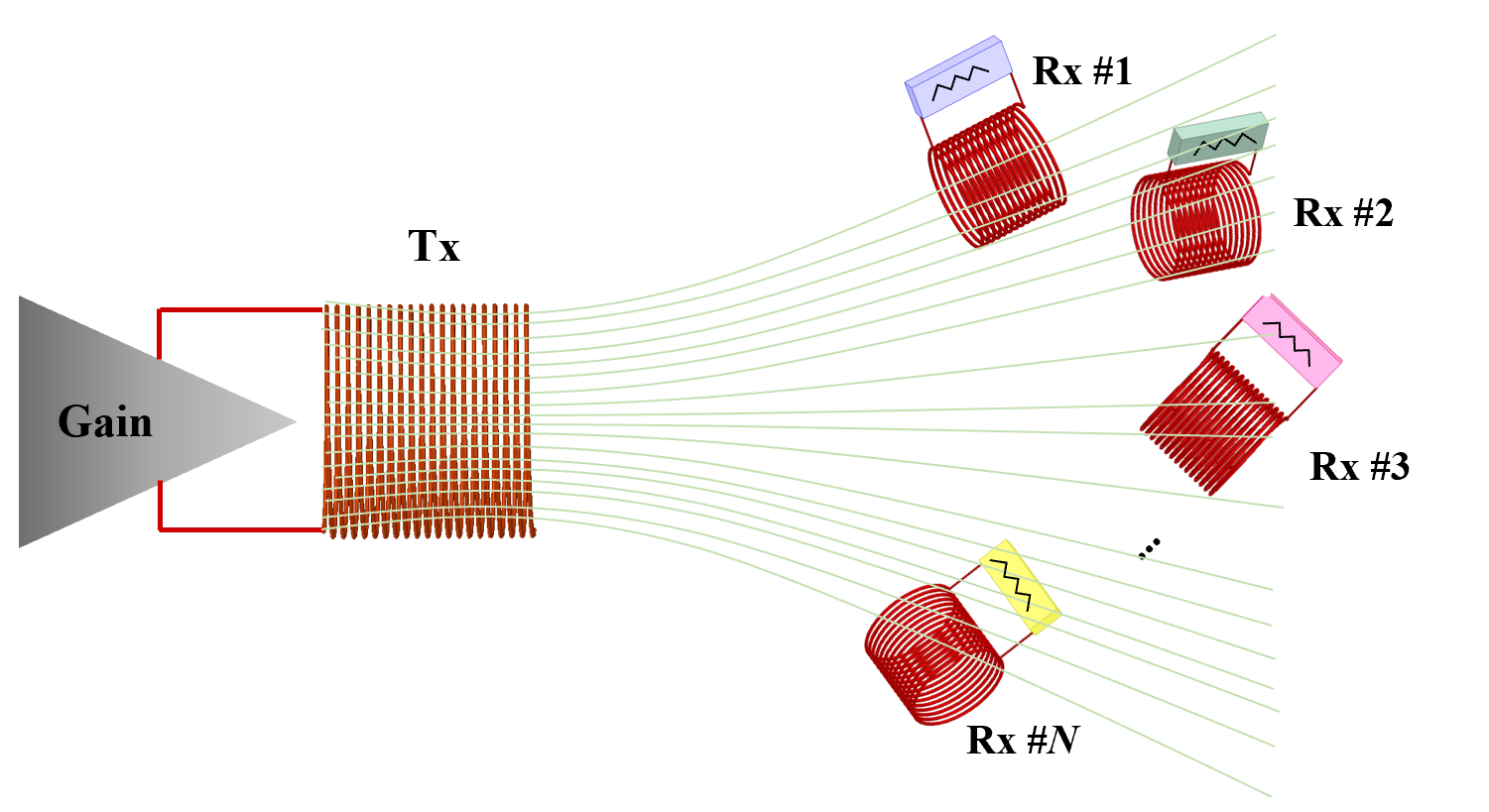}
\caption{Wireless power transfer from one transmitter (Tx), with saturable nonlinear gain,  to $N$ inductively coupled receivers (Rx). The system exhibits an exceptional point of degeneracy that separates the strong from the weak coupling regime.}
\label{fig:1}
\end{figure}

Previous studies proved the potential of WPT using a magnetic inductive link \cite{kurs2007wireless,karalis2008efficient,kiani2011design,ho2013midfield}. Two magnetically coupled resonators, one on the source side and one on the reception side, make up a primary near-field wireless power transmission system. The rates at which energy is injected into and taken out of each resonator and the frequency of the source resonator are all carefully adjusted to ensure efficient power transfer. However, since the change in distance and the relative positioning of the transmitter and receiver affect the coupling factor, the efficiency is out of control in practical scenarios. Therefore, reaching robust WPT without the need for active tuning to get the optimal efficiency is a challenge in designing systems that require dynamic charging. One of the effective methods to solve this problem is using parity-time (PT) symmetric circuits with a nonlinear gain in one resonator, which could balance the loss on the other one \cite{schindler2011experimental,heiss2012physics}. Assawaworrarit et al. \cite{assawaworrarit2017robust} have demonstrated that in WPT links with one receiver possessing PT-symmetry conditions with nonlinear saturation, the system could self-select the operation frequency with optimal efficiency. The PT-symmetry condition can be achieved through a composite gain/loss system which is invariant under the parity (P) and time-reversal (T) operators. The PT-symmetric system made of two resonators has two phases: (i) the \textit{unbroken phase} in which the frequency spectrum is real, and the energy is stored equally between gain and loss regions; and (ii) the \textit{broken phase} in which the eigenfrequencies are complex; therefore, while one mode is growing exponentially, the other one decreases \cite{stehmann2004observation,el2007theory,hodaei2014parity,feng2014single,sakhdari2017pt,chen2018generalized}. In some systems, these two regions are separated by an exceptional point of degeneracy (EPD) \cite{heiss2004exceptional}. An EPD is a special point in a system parameter space at which two or more eigenmodes coalesce in both their eigenvalues and eigenvectors into a single degenerate eigenmode by varying frequency or other parameters of the system \cite{vishik1960solution,lancaster1964eigenvalues,kato1966perturbation,seyranian1993sensitivity,bender1998real}. The main feature of an exceptional point is the strong full degeneracy of the relevant eigenmodes, justifying the presence of "D" in EPD, which stands for "degeneracy" \cite{berry2004physics}.

Compared to previous studies on WPT \cite{xu2016self,ra2018site,liu2019pulsed,feng2020injection}, this work examines the effect of multiple receivers. An additional third coil to improve WPT systems performance has been studied in \cite{wu2022generalized, hao2023frequency, guo2024level,ardila2023analysis}, but these studies deal with systems that are different from ours since they only had one receiving load.
%Indeed, efficient and stable WPT in multiple-transmitter one-receiver systems can be achieved without requiring generally PT-symmetric systems, as also recently discussed in Refs. \cite{wu2022generalized, hao2023frequency, guo2024level}. Also, three coils in a WPT system are used in a way where the additional coil is between the transmitter and receiver to extend their distance range \cite{ardila2023analysis}.
In our work, we also show that the presence of an EPD is crucial for distinguishing between strong and weak coupling regimes. In previous works, this separation is typically explained in terms of coupling factors and circuit parameters. However, by leveraging the underlying physics of EPDs, we provide a deeper interpretation of this transition, revealing additional physical insights. Notably, beyond the EPD point, the system operates in the strong coupling regime, achieving high power transfer efficiency and reliably delivering power to each receiver. Additionally, our design offers greater flexibility in selecting circuit parameters compared to \cite{feng2020injection}. The only requirement is that the natural resonance frequencies are matched before coupling. Once this condition is met, all circuit elements in each resonator can be adjusted accordingly.

In this paper, we first generalize the concept of EPDs in WPT systems with multiple receivers using coupled-mode theory without resorting to PT symmetry. Figure \ref{fig:1} illustrates the general approach of using inductive coupling for power transfer to multiple receivers. By leveraging EPD characteristics, we establish a nearly constant efficiency region in a non-PT-symmetric system for power transfer to {\em multiple} receivers, even when the resonators and their couplings to the main transmitter are not identical. This approach eliminates the need for PT symmetry, requiring only the presence of an EPD to delineate strong and weak coupling regions. Additionally, a saturable nonlinear gain element is employed in the transmitter, enabling a steady-state oscillation in any regime. Finally, we theoretically and experimentally validate this method in a WPT system with one transmitter and two receivers, though the proposed framework extends to systems with any number of receivers.

\section{EPD Condition for Wireless Power Transfer to Multiple Receivers}
The system in Fig. \ref{fig:1} with $N$ receivers is described using coupled-mode theory \cite{haus1984optoelectronics, hodaei2014parity, assawaworrarit2017robust} as
\begin{equation} 
\begin{split}
\dot{a}_{m}(t) & =\left(i \omega_{m}-\gamma_{m}+g_{g a i n} \delta_{m 0}\right) a_{m}(t) \\
& -\sum_{n \neq m} i K_{m n} a_{n}(t)+s_{m}(t).
\end{split}
\label{eq:CMT}
\end{equation}
where $a_{m}$ represents the complex mode amplitude in the $m$-th resonator, defined such that $\left|a_{m}\right|^{2}$ corresponds to the energy stored in that resonator, with $m=0,1,2, \ldots, N$.
The term $\omega_m$ represents the angular frequencies of the uncoupled (i.e., isolated) $m$-th resonator with loss factor $\gamma_{m}$. The coupling coefficient between the $m$-th and the $n$-th resonator is defined by $K_{m n}$. The transmitter is denoted by $m=0$, whereas the $N$ receivers are denoted by $m=1,2,3, \ldots, N$.
In Eq. (\ref{eq:CMT}), the gain in the transmitter is multiplied by the Kronecker delta $\delta_{m 0}$ that is equal to 1 for $m=0$ and 0 otherwise. We assume that the time evolution of all quantities are $a_{m} \propto e^{i \omega t}$ where $\omega$ is the considered angular frequency. The term $s_{m}$ is a possible source in each resonator, though in our case we assume that there are no external sources, as we look at the eigenfrequencies of the system and the self-oscillatory regime induced by the presence of gain.

\begin{figure}[t]
\centering	
\includegraphics[width=1\columnwidth]{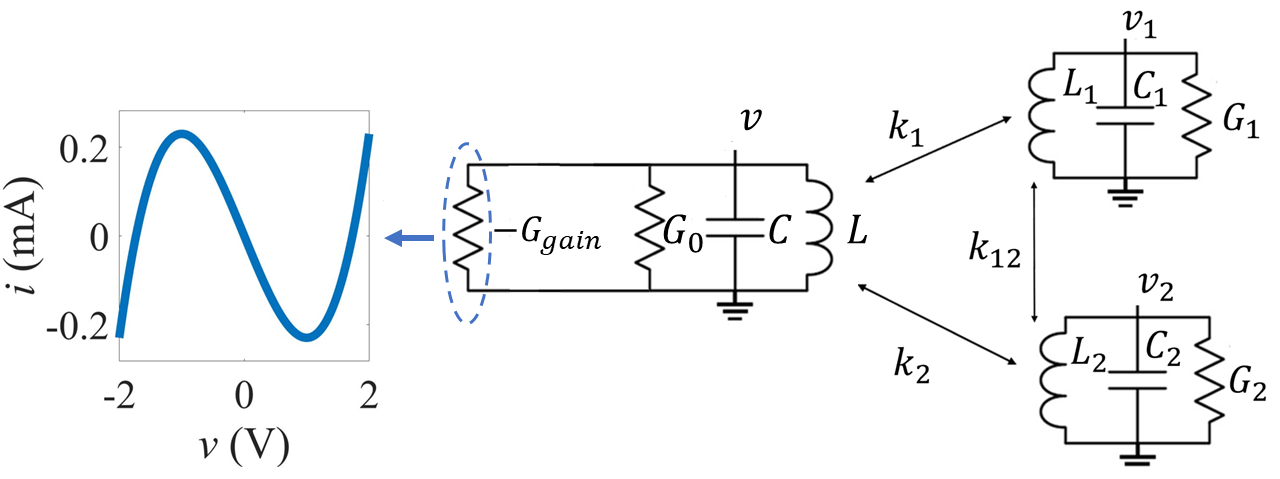}
\caption{Self-oscillating wireless power transfer scheme utilizing three coupled resonators. The transmitter is powered by a nonlinear gain element $(-G_{\text {gain}})$, characterized by the depicted $i-v$ curve. The two receiver resonators are each terminated with linear conductances $G_{1}$ and $G_{2}$, respectively.}
\label{fig:2}
\end{figure}

To simplify the notation and the experimental demonstration, in the rest of the paper, we focus on the WPT case with one transmitter and two receivers, with inductive magnetic couplings as shown in Fig. \ref{fig:2}. The coupling factors between the transmitter and two receivers are denoted as $K_{01}$ and $K_{02}$, but for the sake of brevity, we instead use $K_{1}$ and $K_{2}$, respectively. The effective gain in the transmitter is $-g = -g_{\text{gain}} + \gamma_{0}$, where $g_{\text{gain}}$ is realized using electronic components, and the term $\gamma_{0}$ represents the intrinsic loss rate in the transmitter, mainly occurring in the inductor. Based on the coupled-mode theory for the schematic in Fig. \ref{fig:2}, the system dynamics are described by the following equations
\begin{equation}
\frac{d}{d t}\left[\begin{array}{c}
a \\
a_1 \\
a_2
\end{array}\right]=\left[\begin{array}{ccc}
i \omega_0+g & -i K_1 & -i K_2 \\
-i K_1 & i \omega_1-\gamma_1 & -i K_{12} \\
-i K_2 & -i K_{12} & i \omega_2-\gamma_2
\end{array}\right]\left[\begin{array}{c}
a \\
a_1 \\
a_2
\end{array}\right].
\label{eq:ODE_system}
\end{equation}
For simplicity, we assume that the loss rates of the receivers are equal, i.e., $\gamma_{1}=\gamma_{2}=\gamma_{\mathrm{r}}$. Additionally, we neglect the effect of mutual coupling between the two receivers $\left(K_{12}=0\right)$ since their inductors are pretty small and necessarily not positioned close to each other in practical scenarios. Furthermore, we assume that the resonators in the transmitter and receivers have the same natural frequency when uncoupled ($\omega_{0}=\omega_{1}=\omega_{2}$). To find the eigenfrequencies of the system, we need to solve the eigenvalue problem associated with the matrix in Eq. \eqref{eq:ODE_system}. This involves finding the roots of the characteristic equation, 
\begin{equation}
\begin{split}
       (\omega &- \omega_0 - i\gamma_{\mathrm{r}} ) \\ &\left[\left(\omega -\omega_0  + i g\right)\left(\omega - \omega_0 - i\gamma_{\mathrm{r}} \right) - (K_1^2 + K_2^2)\right] = 0.
\end{split}
\end{equation}
The system has three eigenfrequencies,
\begin{equation}
   \omega = 
     \begin{cases}
       \omega_{0}+i \gamma_{\mathrm{r}}\\
        \omega_{0}+\frac{1}{2}\left( i(\gamma_{\mathrm{r}}-g) - \omega_{\Delta} \right) \\
        \omega_{0}+\frac{1}{2}\left( i(\gamma_{\mathrm{r}}-g) + \omega_{\Delta}\right)
     \end{cases}
     \label{eq:eigenfrequency},
\end{equation}
where $\omega_{\Delta} = \sqrt{4(K_{1}^{2}+K_{2}^{2})-(g+\gamma_{\mathrm{r}})^2}$. The corresponding eigenvectors of the system, after a proper dimensional normalization, are

\begin{equation}
\begin{split}
\mathbf{\Psi} = \left[\begin{array}{c} 0 \\
-K_2\\
K_1
\end{array}\right],
\left[\begin{array}{c}  i\frac{\gamma_{\mathrm{r}}+g}{2} + \frac{\omega_{\Delta}}{2} \\
K_1\\
K_2\end{array}\right],
\left[\begin{array}{c}  i\frac{\gamma_{\mathrm{r}}+g}{2} - \frac{\omega_{\Delta}}{2} \\
K_1\\
K_2\end{array}\right].
\end{split}
\label{eq:eigenvector}
\end{equation}

The first solution in Eq. (\ref{eq:eigenfrequency}) represents a decaying resonant mode. The other two are real-valued eigenfrequency solutions given that the gain conductance is such that $g=\gamma_{\mathrm{r}}$ and  $\omega_{\Delta}$ is real, i.e., for strong coupling between the transmitter and the two receivers that is defined by $\left(K_1^2+K_2^2 \right)> \gamma^2_{\mathrm{r}}$. Furthermore, two of the three eigenfrequencies are degenerate and real when

\begin{equation}
\gamma_{\mathrm{r}}=\sqrt{K_{2}^{2}+K_{1}^{2}}.
\label{eq:epd}
\end{equation}

\begin{figure}
\centering	
\includegraphics[width=1\columnwidth]{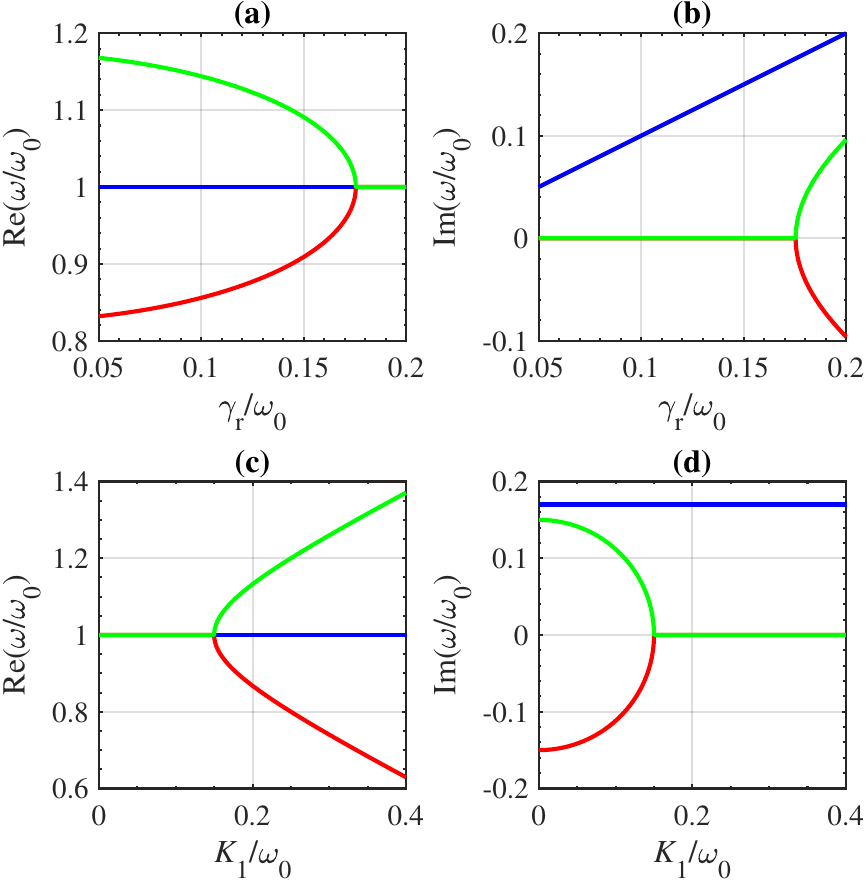}
\caption{(a) Real and (b) imaginary parts of the angular eigenfrequencies for the system in Fig. \ref{fig:2} by varying the receiver loss rate $\gamma_{\mathrm{r}}$, assuming $g=\gamma_{\mathrm{r}}$. The EPD of order two (the bifurcation point) happens at $\gamma_{\mathrm{r}} / \omega_{0}=0.17$, with $K_{1} / \omega_{0}=$ $K_{2} / \omega_{0}=0.124$, where the eigenvalues of the system coalesce. (c) Real and (d) imaginary parts of the angular eigenfrequencies by varying the coupling rate to receiver \#1 while the coupling rate to receiver \#2 is kept constant at $K_{2} / \omega_{0}=0.08$ and $\gamma_{\mathrm{r}} / \omega_{0}=0.17$. The EPD occurs at $K_{1, \mathrm{e}} / \omega_{0}=0.15$.}
\label{fig:3}
\end{figure}

Looking at the eigenvectors (\ref{eq:eigenvector}), we observe that Eq. (\ref{eq:epd}) defines a second-order EPD at $\omega_{\mathrm{e}}=\omega_{0}$. Around this EPD, the two eigenvalues exhibit the typical square root-like (second-order EPD) behavior when the system is perturbed, leading to the typical bifurcation diagrams in Fig. \ref{fig:3}. There, we vary the loss factor in the receivers, and the coupling factor $K_1$ with the first receiver. To better understand the instability of the circuit, we plot the eigenfrequency branches in different colors.  As shown in Fig. \ref{fig:3}, two eigenfrequencies are complex-valued away from the EPD point when $\sqrt{K_{2}^{2}+K_{1}^{2}}<\gamma_{\mathrm{r}}$, i.e., in the so-called weak coupling regime. An eigenfrequency with a negative imaginary part (red branch in Fig. \ref{fig:3}) indicates an exponentially growing signal, and the oscillation frequency is associated with the real part of the eigenfrequency. Instead, in the strong coupling regime, defined by  $\sqrt{K_{2}^{2}+K_{1}^{2}}> \gamma_{\mathrm{r}}$, there are two real-valued eigenfrequencies. The EPD point given by Eq. (\ref{eq:epd}) separates these two regimes. It is also interesting to observe that the system's dynamics are based on the overall combined coupling 
 $\sqrt{K_{2}^{2}+K_{1}^{2}}$ and not on the individual ones.

We also observe another important regime: one of the eigenfrequencies in Eq. (\ref{eq:eigenfrequency}) is always real when  the gain conductance is chosen as \cite{zhou2018nonlinear}
\begin{equation}
g= \min \{\gamma_{\mathrm{r}},\:\left(K_1^2+K_2^2 \right)/\gamma_{\mathrm{r}}\},
\label{eq:gsaturated}
\end{equation}
while the other two solutions have a positive imaginary part, leading the signal to decay over time. We adopt this gain model in Appendix \ref{App: B} for the stability analysis of the self-oscillating WPT system showing that this is the gain value at which the system saturates after reaching a steady-state, i.e., saturated, regime. Indeed, under this regime, the system oscillates without decaying or growing exponentially, hence implying that the oscillation frequency is real-valued. The two gain regimes, $g=\gamma_r<\sqrt{K_{2}^{2}+K_{1}^{2}}$ and $g=\left(K_1^2+K_2^2 \right)/\gamma_{\mathrm{r}}< \gamma_{\mathrm{r}}$  are separated by $g=\gamma_{\mathrm{r}}=\sqrt{K_{2}^{2}+K_{1}^{2}}$ that is the EPD condition in Eq. (\ref{eq:epd}), implying that the analysis of the EPD is important to understand the two regimes of operations. 

\begin{figure}
\centering	
\includegraphics[width=1\columnwidth]{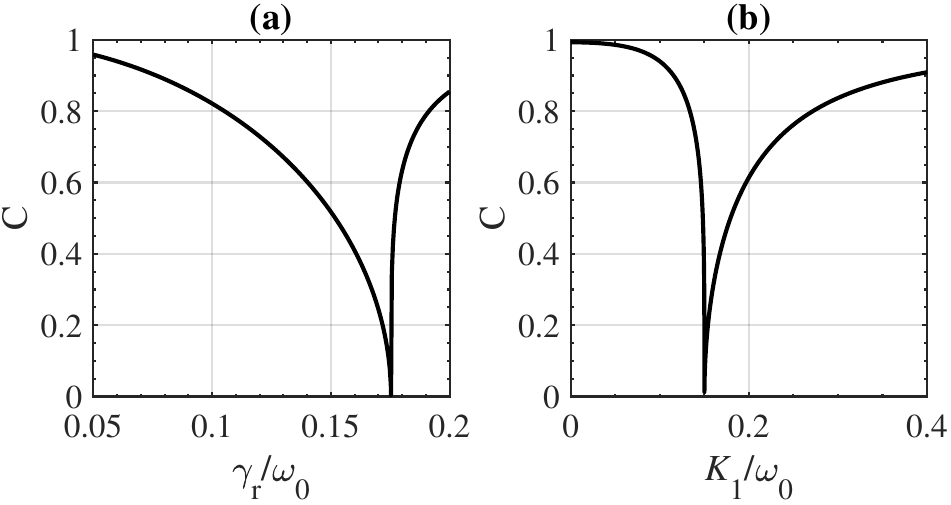}
\caption{Coalescence parameter of two eigenvectors by varying parameters, in the strong and weak coupling regimes separated by the EPD (the null). (a) Varying the receivers loss rate $\gamma_{\mathrm{r}}$, assuming $g=\gamma_{\mathrm{r}}$. The second-order EPD occurs at $\gamma_{\mathrm{r}} / \omega_{0}=0.17$, with $K_{1} / \omega_{0}=$ $K_{2} / \omega_{0}=0.124$, where the eigenvalues of the system coalesce. (b) Varying the coupling rate to receiver \#1, while the coupling rate to receiver \#2 is kept constant at $K_{2} / \omega_{0}=0.08$ and $\gamma_{\mathrm{r}} / \omega_{0}=0.17$. The EPD occurs at $K_{1, \mathrm{e}} / \omega_{0}=0.15$.}
\label{fig:4}
\end{figure}

The coalescence of at least two eigenvalues {\em and} eigenvectors defines the EPD that occurs when $\omega_{\Delta}=0$. To analyze how the eigenvectors coalesce, we use the coalescence parameter tool presented in \cite{abdelshafy2019exceptional,nikzamir2022highly}. While the coalescence of eigenvalues is required for an EPD, the coalescence of eigenvectors ensures its existence.  To calculate how close the system is to a second-order degeneracy condition at the frequency of interest, we calculate the coalescence parameter (C) as 
\begin{equation}\mathrm{C}=|\mathrm{sin}\theta_{mn}|,\quad \cos\theta_{mn}=\frac{\left|\left\langle \mathbf{\mathbf{\Psi}}_{m},\mathbf{\mathbf{\Psi}}_{n}\right\rangle \right|}{\left\Vert \mathbf{\mathbf{\Psi}}_{m}\right\Vert \left\Vert \mathbf{\mathbf{\Psi}}_{n}\right\Vert },
\label{eq:coalescence}
\end{equation}
where $\theta_{mn}$, with $m=2$ and $n=3$, represents the angle between the second and third eigenvectors given in Eq. \eqref{eq:eigenvector}, as explained earlier, in a three-dimensional complex vector space via the inner product $\left\langle \mathbf{\Psi}_{m},\mathbf{\Psi}_{n}\right\rangle =\mathbf{\Psi}_{m}^{\dagger}\mathbf{\Psi}_{n}$, where the dagger symbol $\dagger$ denotes the complex conjugate transpose operation. The absolute value is represented by $\left|\right|$, and the norm of a complex vector is given by $\left\Vert \mathbf{\Psi}\right\Vert =\sqrt{\left\langle \mathbf{\Psi},\mathbf{\Psi}\right\rangle}$. Figure \ref{fig:4} illustrates the coalescence of the eigenvectors, highlighting the degeneracy of the modes in the system at the EPD point. This EPD represents a separation between the strong and weak coupling regimes.

In this paper, the values of the inductance $(L)$ and capacitance $(C)$ are chosen in a way to have matched resonance frequencies in receivers, though an EPD would exist also when the receivers have different resonant frequencies.

\section{Theoretical Wireless Power Transfer Efficiency}
To analyze the total power transferred to the loads of both receivers, the total power efficiency $\eta_{\mathrm{t}}$ is calculated for different ranges of coupling using coupled-mode theory \cite{kurs2010simultaneous,assawaworrarit2017robust} as
\begin{equation}
\begin{split}
\eta_{\mathrm{t}}&=\frac{\gamma_{1}\left|a_{1}\right|^{2}+\gamma_{2}\left|a_{2}\right|^{2}}{\gamma_{0}|a|^{2}+\gamma_{1}\left|a_{1}\right|^{2}+\gamma_{2}\left|a_{2}\right|^{2}}\\
& =    \begin{cases}
       \frac{\gamma_{\mathrm{r}}}{\gamma_{\mathrm{r}}+\gamma_{0}}, &\quad\gamma_{\mathrm{r}} \leq \sqrt{K_{2}^{2}+K_{1}^{2}}\\
       \frac{K_{1}^{2}+K_{2}^{2}}{K_{1}^{2}+K_{2}^{2}+\gamma_{\mathrm{r}} \gamma_{0}}, &\quad\gamma_{\mathrm{r}}>\sqrt{K_{2}^{2}+K_{1}^{2}} \\
     \end{cases}
\end{split}
\label{eq:eta_t}
\end{equation}

\begin{figure}
\centering	
\includegraphics[width=1\columnwidth]{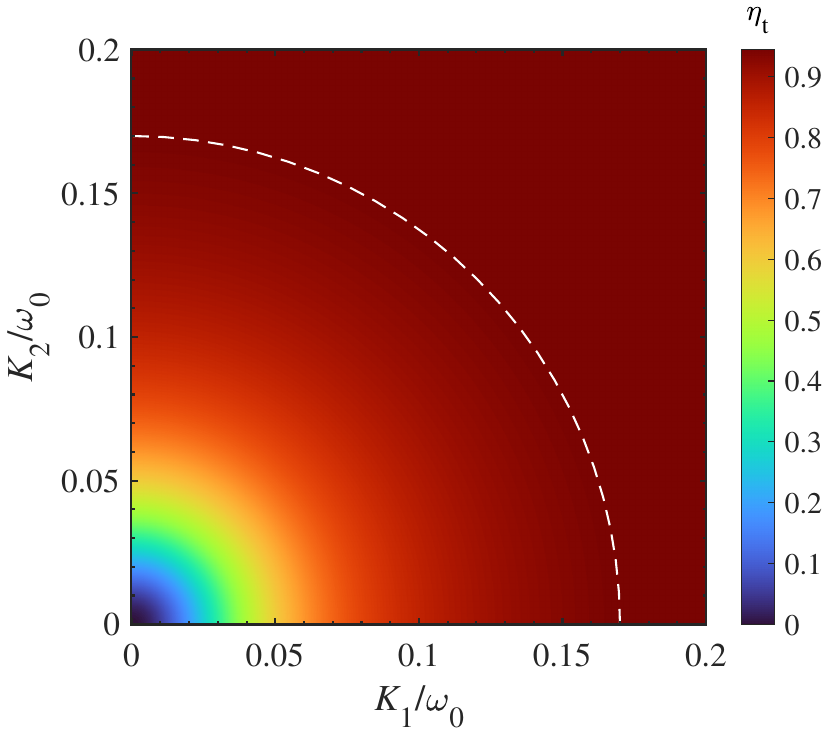}
\caption{The efficiency of total power transferred to the receivers by varying coupling factors $K_1/\omega_0$ and $K_2/\omega_0$. The weak (inner) and strong (outer) coupling regions are separated by the dashed-white circle. The dashed-white circle represents the EPD condition $K_{2}^{2}+K_{1}^{2}=$ $\gamma_{\mathrm{r}}^{2}$, which defines the boundary between the two coupling regimes. The values of $\gamma_{\mathrm{r}} / \omega_{0}=0.17$ and $\gamma_{0} / \omega_{0}=0.001$ are used in this analysis.}
\label{fig:5}
\end{figure}

Details are in Appendix \ref{App: A}. In the strong coupling regime, the total efficiency remains independent of the coupling factors and approaches unity when the intrinsic loss rate in the transmitter ($ \gamma_0$) is negligible. Since there are two coupling factors that can vary, a two-dimensional strong coupling region exists, as shown in Fig. \ref{fig:5} (outside the white-dashed circle with radius $\gamma_{\mathrm{r}} / \omega_{0}$), resulting in high power transfer efficiency. This means that in the strong coupling regime (outside the white-dashed circle), the total power delivered to both receivers remains constant, demonstrating the system’s robustness in efficiently delivering power. On the other hand, in the weak coupling regime, variations in the coupling factors affect the total power efficiency of both receivers.

To analyze the power transferred to each resonator individually, the coupling rate to the other resonator is assumed to be fixed. For instance, assuming $K_{2}$ is constant and $K_{2}<\gamma_{\mathrm{r}}$, we define $K_{1, \mathrm{e}}=\sqrt{\gamma_{\mathrm{r}}^{2}-K_{2}^{2}}$ as the EPD condition for different coupling regions in receiver \#1, based on Eq. \eqref{eq:epd}. The strong coupling regime $K_{1} \geq K_{1, e}$ is now explicitly defined for receiver \#1, and the same calculation process is valid for the second receiver. Assuming $K_{2}$ remains constant, the efficiency for receiver \#1 is expressed as
\begin{equation}
\begin{split}
\eta_{1}&=\frac{\gamma_{1}\left|a_{1}\right|^{2}}{\gamma_{0}|a|^{2}+\gamma_{1}\left|a_{1}\right|^{2}+\gamma_{2}\left|a_{2}\right|^{2}} \\
&=     \begin{cases}
       \frac{\gamma_{\mathrm{r}}}{\gamma_{\mathrm{r}}+\gamma_{0}} \frac{K_{1}^{2}}{K_{1}^{2}+K_{2}^{2}}, &\quad K_{1} \geq K_{1, e}\\
       \frac{K_{1}^{2}}{K_{1}^{2}+K_{2}^{2}+\gamma_{\mathrm{r}} \gamma_{0}}, &\quad K_{1}<K_{1, e} \\
     \end{cases}
\end{split}
\label{eq:eta1}
\end{equation}

Assuming $K_{2}<\gamma_{\mathrm{r}}$, in the strong coupling regime where $K_{1} \geq K_{1, \mathrm{e}}$, the power delivered to receiver \#1 remains roughly constant, even as $K_{1}$ changes, as long as $K_{1} \gg K_{2}$. This means that, for a range of $K_1$ values, the power delivered to receiver \#1 doesn’t vary significantly with its coupling factor to the transmitter. Similarly to what was mentioned for the total power delivered to both receivers, if receiver \#1 is in the strong coupling regime and receiver \#2 is far away, the power efficiency approaches unity. This indicates that, as long as transmitter loss is negligible, all of the power generated by the transmitter is delivered to receiver \#1. In contrast, in the weak coupling region for receiver \#1, where $K_{1}<K_{1, e}$, variations in the coupling factor have a more pronounced impact on its efficiency. 

However, when $K_{2}>\gamma_{\mathrm{r}}$, the operating points for both receivers lie outside the white dashed circle in Fig. \ref{fig:5}. Since both receivers are in the strong coupling regime, both $K_{1}$ and $K_{2}$ have a significant impact on the individual efficiency. In this case, varying $K_{1}$  does not lead to an EPD, and the efficiency for receiver \#1 does not level to a constant value. This means that the system's robustness is not as strong as expected, but the power delivered to each receiver can still be directly controlled by its coupling factor, assuming the transmitter’s losses are negligible. The same analysis is valid for receiver \#2 when fixed $K_{1}$ is chosen such that $K_{1}>\gamma_{\mathrm{r}}$. This highlights the importance of achieving an EPD for the robustness of power transfer in the system, which has been a key focus throughout this paper. Moreover, based on the operating point positions in Fig. \ref{fig:5}, the total efficiency, defined as the sum of the individual efficiencies, $\eta_{\mathrm{t}}=\eta_{1}+\eta_{2}$, yields Eq. \eqref{eq:eta_t}.

\section{Oscillatory Saturated Regime}
The negative conductance feeds power to the transmitter resonator, which is coupled through a magnetic link to the receivers with loads denoted by $G_{1}$ and $G_{2}$, as shown in Fig. \ref{fig:2}. The transient behavior of the system is obtained by using the time-domain (TD) circuit simulator Keysight ADS, with the initial condition at the transmitter $v(t=0)=1\:\mathrm{mV}$. 
The circuit parameters in the Tx part are $C=330\:\mathrm{pF}, L=84.45\:\mu \mathrm{H}, G_{\text {gain }}=$ $0.345\:\mathrm{mS}, G_{0}=0.002\: \mathrm{mS}$, and the Rx circuit parameters are $C_{1}=$ $C_{2}=3.30\:\mathrm{nF}, L_{1}=L_{2}=8.445\:\mu \mathrm{H}, G_{1}=G_{2}=3.45\:\mathrm{mS}$. 
Therefore, since $\omega_{m}= 1/\sqrt{L_{m} C_{m}}$, we study the case with $\omega_0=\omega_1=\omega_2$ in the remainder of the paper. The inductance and capacitance values on each resonator are selected to ensure matched resonance frequencies, which in the experiment is achieved through capacitor tuning. 
The inductive mutual coupling coefficient between the $m$-th and the $n$-th resonator is $k_{m n}=M_{m n} / \sqrt{L_{m} L_{n}}$, where $M_{m n}$ is their mutual inductance. The transmitter is denoted by $m=0$. Therefore, consistent with our earlier assumptions, we consider the direct coupling between the two receivers to be negligible, i.e., $k_{12}=k_{21} \approx 0$. Hence, the coupling factors between the transmitter and two receivers $k_{01}$ and $k_{02}$ are denoted with $k_{1}$ and $k_{2}$, respectively, for the sake of brevity. 

The relations with the parameters used in coupled-mode theory are  $\gamma_{m}=G_{m} \omega_{m} \sqrt{L_{m} / C_{m}}$. The circuit parameters lead to $\gamma_1=\gamma_2=\gamma_{\mathrm{r}}$ with the value of $\gamma_{\mathrm{r}} / \omega_{0}=0.17$, and $\gamma_{0} / \omega_{0}=0.001$, satisfying the EPD condition discussed in the previous section. Furthermore, we assume that $K_{m}=\omega_{0} k_{m}$.

In the TD simulations, the nonlinear saturable gain element is realized using a cubic model with an $i-v$ curve described as $i=-G_{\text {gain }} v+\alpha v^{3}$, where $-G_{\text {gain }}$ is the negative slope of $i-v$ curve, describing the small-signal negative admittance region, and $\alpha$ is the third-order nonlinearity constant that models the saturation characteristic of the device \cite{oshmarin2021experimental,nikzamir2022highly}. The value of the saturation characteristic $\alpha$ determines the steady-state saturation amplitude and in this paper is set as $\alpha=G_{\text {gain }}/3$. The relation with coupled-mode theory is established with the effective small-signal gain in the transmitter as $-g = -g_{\text{gain}} + \gamma_{0}$, where $g_{\text{gain}} = G_{\text{gain}}\omega_{0}\sqrt{L/C}$.

\begin{figure}
\centering	
\includegraphics[width=1\columnwidth]{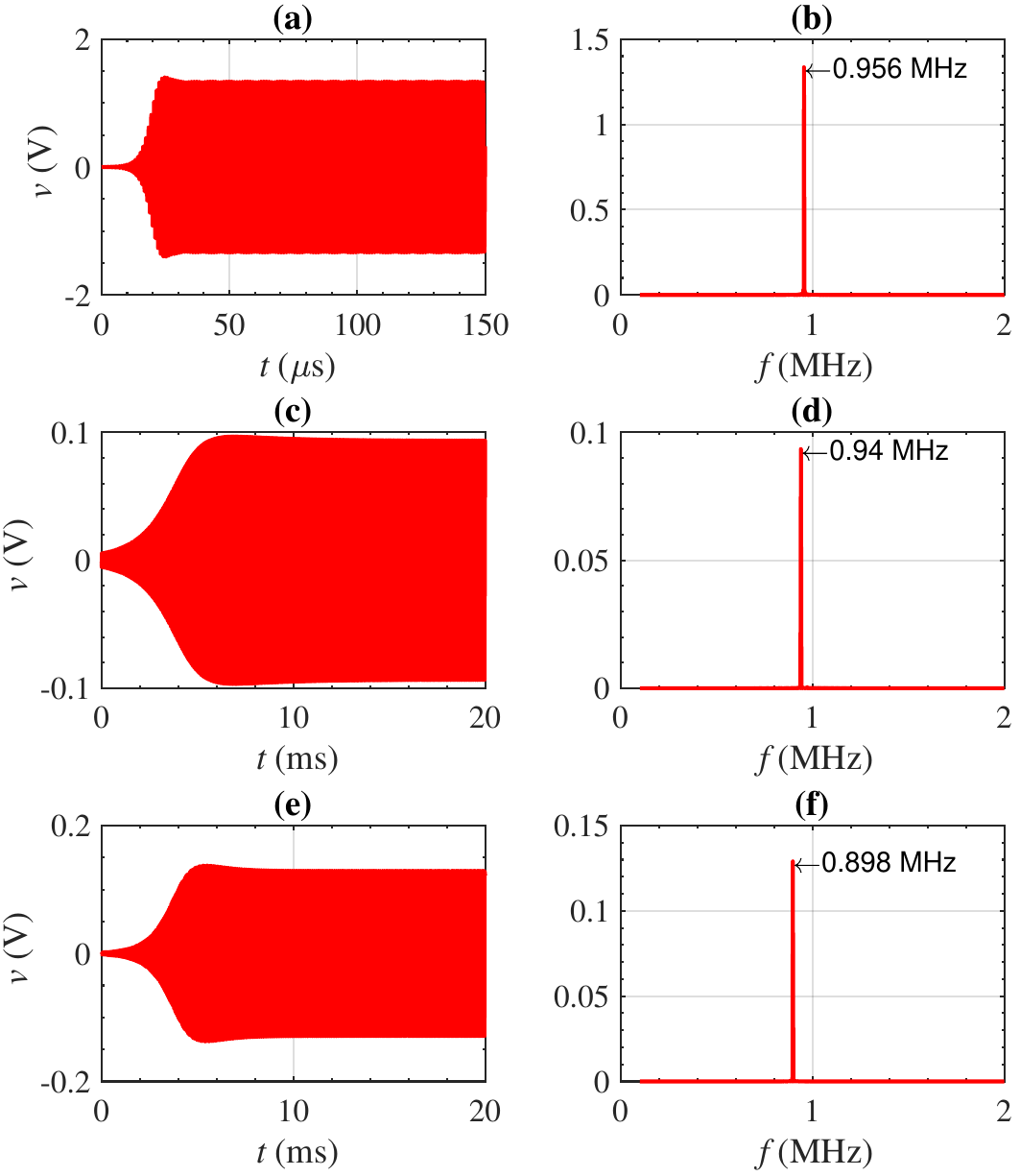}
\caption{Time-domain signal and frequency spectrum of $v(t)$ in the transmitter after reaching saturation due to the nonlinear effects of the active element, showing a single oscillation frequency. (a)(b) System operating away from the EPD in the weak coupling regime, with $k_{1}=0.1$ and $k_{2}=0.08$. The system oscillates at a single frequency of $f= 0.956 \:\mathrm{MHz}$. (c)(d) System operating at the EPD, with the oscillation frequency of $f= 0.94\:\mathrm{MHz}$. (e)(f) System operating away from the EPD in the strong coupling regime, with $k_{1}=0.2$ and $k_{2}=0.08$. The system oscillates at a single frequency of $f= 0.898 \:\mathrm{MHz}$.}
\label{fig:6}
\end{figure}

In the small-signal regime, the system has three modes, but due to the nonlinear gain in the transmitter, the mode that requires the least gain will grow to attain its steady state and saturate out the gain, preventing other modes from gaining access to the gain required for steady-state oscillation. 
Figure \ref{fig:6}(a)(b) shows the simulated TD voltage of the transmitter when operating away from EPD, with $k_{1}=0.1$ and $k_{2}=0.08$, i.e., in the weak coupling regime. The frequency spectrum of the transmitter voltage, after reaching saturation, shows a fundamental operating frequency of oscillation at $f_{\text{osc, weak}}=0.956\:\mathrm{MHz}$, which is close to the linear-regime oscillation frequency shown in Fig. \ref{fig:3}(c), given by $\omega_0/(2\pi) = 1/\left(2\pi\sqrt{LC}\right) = 0.953\:\mathrm{MHz}$. The small difference comes from using nonlinear gain employed in the TD simulation.
Figure \ref{fig:6}(c)(d) shows the TD circuit simulation of the capacitor voltage $v(t)$ in the transmitter when $k_{1}=k_{2}=0.124$, i.e., when the system is supposed to operate at the EPD. The frequency spectrum of the transmitter voltage, after reaching saturation, shows a fundamental operating frequency of oscillation $f_{\text{osc, EPD}}=0.940\:\mathrm{MHz}$, close to the linear-regime oscillation frequency $\omega_0/(2\pi)=0.953\:\mathrm{MHz}$  shown in Fig. \ref{fig:3}(c). One should note that the oscillation frequency in the weak coupling regime is closer to the analytical value in the linear regime $\left(\omega_0/(2\pi)\right)$ than to the oscillation frequency at the EPD. This is because the circuit becomes highly sensitive when operating at an EPD, making the effects of the nonlinear elements on the oscillation frequency more pronounced.

Figure \ref{fig:6}(e)(f) demonstrates the transmitter's TD signal in simulation, away from EPD, with $k_{1}=0.2$ and $k_{2}=0.08$, i.e., in the strong coupling regime. After reaching saturation, the self-oscillating frequency for this case is $f_{\text{osc, strong}}=0.898 \mathrm{MHz}$, based on the FFT of the TD signal. Note that in the linear strongly coupled regime (i.e., for small signal) there are two real eigenfrequencies as shown in Fig. \ref{fig:3}(c). However, the spectrum of the signal of the nonlinear simulation shows only one frequency of oscillation. This happens because the nonlinear gain slightly shifts the system away from the ideal condition, causing one resonance to dominate due to its larger instability during the saturation process.
For TD simulations at the EPD and in the strong coupling regime, the small-signal gain conductance was assumed to be $1\%$ larger than losses in the receivers, i.e. $G_{\mathrm{gain}} = 1.01 G_1 = 1.01 G_2$. This small excess in gain is necessary to make the system slightly unstable to trigger oscillations in these regimes. This is not necessary in the weak coupling regime because the system is inherently unstable due to the negative imaginary part of the eigenfrequency, as shown in Fig. \ref{fig:3} (c). 
For all simulation results in Fig. \ref{fig:6}, the frequency spectrum of the TD voltage signal $v(t)$ at the transmitter's capacitor, after reaching saturation, is obtained by applying the fast Fourier transform (FFT) of the TD signal. The FFT is calculated using $10^{6}$ samples in the time window from $50\:\mu\mathrm{s}$ to $150\:\mu\mathrm{s}$ for the signal in weak coupling regime, as shown in Fig. \ref{fig:6}(a), and from $10\:\mathrm{ms}$ to $20\:\mathrm{ms}$ for the signal at the EPD and in strong coupling regime, as shown in Fig. \ref{fig:6}(c) and (e).

The experimental verification discussed next is based on the setup shown in Fig. \ref{fig:7}(a). It uses the same conductances and LC parameter values for both the transmitter and receivers as used in the TD circuit simulations.

As a preliminary step, after fabricating the three inductors, we fine-tuned the transmitter's capacitance using a parallel rack of capacitors so that the transmitter's resonance frequency matches that of the two uncoupled Rx resonators. We used various capacitor values in parallel to ensure we achieved the desired transmitted resonance frequency, as illustrated in Appendix \ref {App: C} (Fig. \ref{fig:10}). A resistance trimmer was used to adjust the value of the gain rate to the value of $-G_{\text {gain }}$ in the transmitter, based on the calculated value of the required gain derived from theory for the system to have an EPD. The detail for tuning is explained in Appendix \ref{App: C}. The circuit is then let to saturate and establish steady-state oscillations. We used an oscilloscope to measure the TD voltage response in the resonators and a spectrum analyzer to examine the frequency response. These measurements confirmed that the three coupled-coil system, including the active nonlinear component, operates in proximity of the EPD. Figure \ref{fig:7}(b) shows the measured operating self-oscillation frequency in the configuration with receiver \#1 moving while receiver \#2 is fixed. As it is demonstrated, the operating frequency starts varying when the coupling to receiver \#1 gets close to the EPD value ($k_{1}$ getting close to $k_{1, \mathrm{e}}$), and after that it changes more dramatically in the strong coupling region ($k_{1}>k_{1, e}$). The FFT spectrum from the simulated TD signal is well matched to the experimental spectrum (either obtained from the oscilloscope waveform or the spectrum analyzer) for all coupling values $k_{1}$ in the range of $0$ to $0.3$. There is a transition of the steady-state frequency among the frequency branches around the EPD in the measurement. This transition has little effect on transfer efficiency. The slight difference between simulated and measured oscillation frequency is attributed to (i) minor mistuning of the resonators, (ii) detuning caused by the nonlinear component during saturation, and (iii) differences between the initial value for gain in measurement and theory.

\section{Power Transfer Efficiency}
To investigate experimentally the power efficiency characteristics of the self-oscillating WPT system in both the weak and strong coupling regimes separated by the EPD, we move one of the receivers' coils (e.g., receiver \#1) to determine the transmitter and receivers' power changes, while keeping the other receiver (e.g., receiver \#2) fixed. The distance range between the transmitter and receivers' coils in the test is converted to coupling coefficient values by using the mutual inductance coupling factor versus distance obtained from COMSOL Multiphysics simulations shown in Fig. \ref{fig:9} in Appendix \ref{App: C}. 

Efficiency is defined as the power received by the two receivers to the power generated in the transmitter by the nonlinear active element. The power received by the two receivers is obtained by measuring the voltage peak values $V_i$  on each receiver ($i=1,2$)  and calculating the power as $V_{i}^{2} G / 2$. The experiment was repeated with different distances between receiver \#2 (the stationary one) and the transmitter to investigate the effect of the fixed receiver position on the moving receiver's efficiency. The measurement distance for the coupling range is in the interval of $5-30\:\mathrm{~mm}$.

\begin{figure}
\centering	
\includegraphics[width=1\columnwidth]{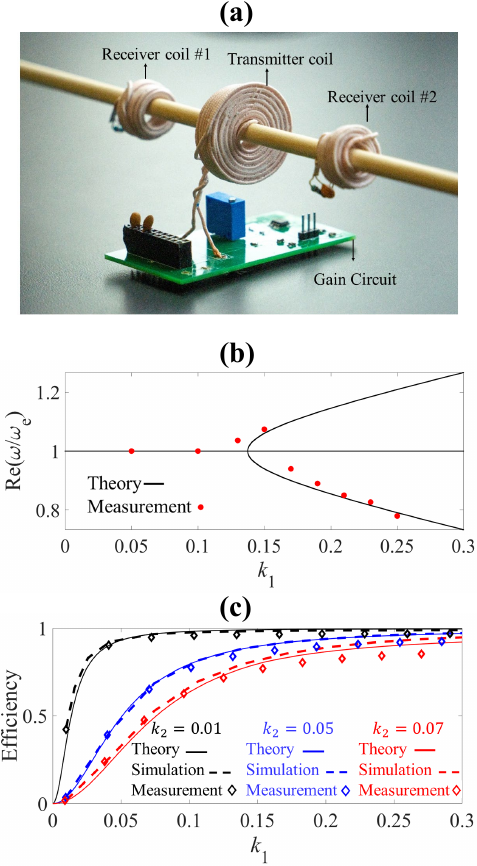}
\caption{(a) The setup used in the measurement of the power transfer efficiency for receiver \#1, as defined in Eq. (\ref{eq:eta1}). (b) Real part of normalized angular eigenfrequencies versus the coupling rate to the receiver \#1, with a fixed coupling value of $k_{2}=0.08$. The theoretical values, based on linear gain, show two real frequencies in the strong coupling regime. The measurement shows the steady state self-oscillation frequency when using saturable nonlinear gain. Only one frequency is observed for every coupling coefficient. (c) Power efficiency for receiver \#1 varying the coupling factor $k_{1}$, shown for three constant $k_{2}$ values: $k_{2}=0.01$ (black), $k_{2}=0.05$ (blue), $k_{2}=0.07$ (red). Solid lines represent the theoretical calculations with linear gain, dashed lines are obtained from time-domain simulations with nonlinear gain, and the diamond symbols indicate experimental measurements.}
\label{fig:7}
\end{figure}

Results in Fig. \ref{fig:7}(c) are calculated as follows: the theory is based on Eq. (\ref{eq:eta1}); simulations and circuit measurements are based on the steady-state saturated regime caused by the nonlinear active gain. The simulation and measurement results agree with the theory from the coupled-mode theory calculation. The results show that the receiver's efficiency after the EPD (i.e., in the strong coupling regime) is almost constant, and approximately independent of the value of the coupling coefficient $k_{1}$. Figure \ref{fig:7}(c) also shows that the value of the second receiver's coupling factor $k_{2}$ affects the maximum efficiency value for receiver \#1. As the second receiver gets closer to the transmitter, $k_{2}$ increases, and the efficiency of the first receiver decreases since the second receiver draws more power from the transmitter. However, the transferred energy to receiver \#1 is still approximately constant over the strong coupling range $k_{1}>k_{1, e}$. Also, the location of receiver \#2 (i.e., the value of the coupling factor $k_{2}$) changes the EPD in the system and consequently redefines the strong-weak coupling region for receiver \#1. This means that when receiver \#2 is located closer to the transmitter (larger $k_2$), the robustness of the power delivered to receiver \#1 is slightly reduced, but the power efficiency does not change significantly. 
The results in Fig. \ref{fig:7}(b)(c), which illustrate the eigenfrequency and efficiency of our system across different coupling regimes, closely resemble those in \cite{feng2020injection} for a single receiver. However, our circuit design provides greater tunability compared to their approach, with the only constraint being the requirement for equal resonance frequencies of the two uncoupled resonators. While maintaining equal loss in both receiver resonators is beneficial, ensuring matched resonance frequencies remains our primary focus.
Notably, our system exhibits self-oscillation at its operating frequency even in the weak coupling regime. Although the power transfer efficiency is not optimal in this regime, the oscillation process is still sustained.

In the theoretical calculations and TD circuit simulations, the intrinsic losses of capacitors and inductors in the receivers were assumed to be negligible compared to losses associated with the loads that are represented by $\gamma_{\mathrm{r}}$ in the coupled-mode theory and $G_1, G_2$ in the TD simulations. 
The slight difference between the value of efficiency in simulation and measurement could be addressed by the presence of losses in the system, such as the intrinsic loss of the coils and the extra loss from the circuit elements and measurement pin joints.

\section{Conclusion}
We have proposed an efficient self-oscillating wireless power transfer method involving one transmitter and multiple receivers exploiting the EPD concept in a non-PT-symmetric structure. This method provides a coupling-independent eigenfrequency region, leading to the design of a system that can transfer power to multiple moving receivers without requiring active tuning of circuit parameters. Variations of both mutual couplings do not affect the robustness of the system because of the saturable nonlinear gain leading to a self-oscillatory regime, with the EPD separating the weak and the strong coupling regimes. Analyses of relatively simple implementation geometries show promising performance characteristics and substantial design tuning is predicted to yield even better results.

A finely tuned EPD point enables a range of constant efficiency without the need for extra electronic circuits in receivers. Moreover, since the receivers and transmitter do not necessarily need to be identical in our proposed method, our method can be useful in a wide range of practical applications, including passive wireless sensing from multiple sites \cite{mohseni2021design}, and powering multiple compact implants and microrobots \cite{hajiaghajani2019patterned}.

\section*{Acknowledgment}
This material is based on work supported by the National Science Foundation under award NSF ECCS-1711975. The authors acknowledge useful discussions with PhD student Benjamin Bradshaw, UC Irvine. 

\appendix
\section{Power Transfer Efficiency}
\label{App: A}
We derive expressions for power transfer efficiency based on the coupled-mode theory equations. Starting with Eq. (\ref{eq:CMT}), for a system made of one transmitter and two receivers, and assuming all quantities $a, a_1$ and $a_2$ include  the complex time-varying factor $e^{i \omega t}$, we have
\begin{gather}
i \omega a =\left(i \omega_{0}+g\right) a-i K_{1} a_{1}-i K_{2} a_{2}, \label{eq:transmitter}\\
i \omega a_{1} =\left(i \omega_{1}-\gamma_{1}\right) a_{1}-i K_{1} a-i K_{12} a_{2}, \\
i \omega a_{2}  =\left(i \omega_{2}-\gamma_{2}\right) a_{2}-i K_{2} a-i K_{12} a_{1}.
\end{gather}
Assuming $K_{12}=0$, the $a_{1}$ and $a_{2}$ mode amplitudes in each receiver resonator are calculated as
\begin{gather}
a_{1}=\frac{-i K_{1}}{i\left(\omega-\omega_{1}\right)+\gamma_{1}} a, \label{eq:a1_weak}\\
a_{2}=\frac{-i K_{2}}{i\left(\omega-\omega_{2}\right)+\gamma_{2}} a. \label{eq:a2_weak}
\end{gather}

In the following, we assume that the natural frequencies of all three resonators (when uncoupled) are matched, i.e., $\omega_{1}=\omega_{2}=$ $\omega_{0}$. For an operating frequency $\omega$ such that $\omega=\omega_{0}$ (which is valid in the weak coupling regime, and only as an approximation in the strong coupling regime), we evaluate $a_{1}$ and $a_{2}$ in simple terms as
\begin{gather}
a_{1}=\frac{-i K_{1}}{\gamma_{1}} a, \\
a_{2}=\frac{-i K_{2}}{\gamma_{2}} a.
\end{gather}
%
%Above, we did not consider the complex frequency solution in Eq. (\ref{eq:eigenfrequency}) because it decays exponentially. 
Now, the efficiency is calculated based on the power delivered to the receivers' loads to the total power generated that is equal to the power dissipated by all three circuits (the total loss is in the receivers' loads plus the intrinsic loss $\gamma_0$ in the transmitter). The total efficiency, assuming $\gamma_{1}=\gamma_{2}=\gamma_{\mathrm{r}}$, in the weak coupling regime is
\begin{equation}
\begin{split}
\eta_{\mathrm{t}} & = \frac{\text {Power transferred to both receievrs}}{\text {Total power generated}} \\
 & =\frac{\gamma_{1}\left|a_{1}\right|^{2}+\gamma_{2}\left|a_{2}\right|^{2}}{\gamma_{0}|a|^{2}+\gamma_{1}\left|a_{1}\right|^{2}+\gamma_{2}\left|a_{2}\right|^{2}} \\
 & =\frac{\gamma_{1}\left|\frac{-iK_{1}}{\gamma_{1}}\right|^{2}|a|^{2}+\gamma_{2}\left|\frac{-iK_{2}}{\gamma_{2}}\right|^{2}|a|^{2}}{\gamma_{0}|a|^{2}+\gamma_{1}\left|\frac{-iK_{1}}{\gamma_{1}}\right|^{2}|a|^{2}+\gamma_{2}\left|\frac{-iK_{2}}{\gamma_{2}}\right|^{2}|a|^{2}} \\
& =\frac{K_{1}^{2}+K_{2}^{2}}{K_{1}^{2}+K_{2}^{2}+\gamma_{\mathrm{r}} \gamma_{0}}.
\end{split}
\end{equation}

The efficiency of each individual receiver can be derived following an analogous procedure. For instance, the efficiency of receiver \#1 in the weak coupling regime is
\begin{equation}
\begin{split}
\eta_{1} &=\frac{\text {Power transferred to receiver}\:\# 1}{\text {Total power generated}} \\
&=\frac{\gamma_{1}\left|a_{1}\right|^{2}}{\gamma_{0}|a|^{2}+\gamma_{1}\left|a_{1}\right|^{2}+\gamma_{2}\left|a_{2}\right|^{2}} \\
&=\frac{\gamma_{1}\left|\frac{-iK_{1}}{\gamma_{1}}\right|^{2}|a|^{2}}{\gamma_{0}|a|^{2}+\gamma_{1}\left|\frac{-iK_{1}}{\gamma_{1}}\right|^{2}|a|^{2}+\gamma_{2}\left|\frac{-iK_{2}}{\gamma_{2}}\right|^{2}|a|^{2}} \\
&=\frac{K_{1}^{2}}{K_{1}^{2}+K_{2}^{2}+\gamma_{\mathrm{r}} \gamma_{0}}.
\end{split}
\end{equation}

However, in the strong coupling regime, we use the precise value of the eigenfrequencies $\omega=\omega_{0} \pm \sqrt{K_{1}^{2}+K_{2}^{2}-\gamma_{\mathrm{r}}{ }^{2}}$, while still assuming $\omega_{1}=\omega_{2}=\omega_{0}$ and $\gamma_{1}=\gamma_{2}=\gamma_{\mathrm{r}}$. With this, and using Eqs. (\ref{eq:a1_weak}) and (\ref{eq:a2_weak}), the expressions for $a_{1}$ and $a_{2}$ in the strong coupling regime are

\begin{gather}
a_{1}=\frac{-i K_{1}}{\pm i \sqrt{K_{1}^{2}+K_{2}^{2}-\gamma_{\mathrm{r}}^{2}}+\gamma_{\mathrm{r}}} a, \\
a_{2}=\frac{-i K_{2}}{\pm i \sqrt{K_{1}^{2}+K_{2}^{2}-\gamma_{\mathrm{r}}^{2}}+\gamma_{\mathrm{r}}} a.
\end{gather}
The magnitudes of $a_{1}$ and $a_{2}$ are
\begin{gather}
\left|a_{1}\right|=\frac{K_{1}}{\sqrt{K_{1}^{2}+K_{2}^{2}}}|a| \label{eq:abs_a1_storng}, \\
\left|a_{2}\right|=\frac{K_{2}}{\sqrt{K_{1}^{2}+K_{2}^{2}}}|a| \label{eq:abs_a2_storng}.
\end{gather}
However, in reality, only the solution with the $-$ sign is the one that matters because it is the one associated with the saturated gain, though the sign does not affect the power result. 

Substituting Eqs. (\ref{eq:abs_a1_storng}) and (\ref{eq:abs_a2_storng}) in the efficiency expression, the total efficiency in the strong coupling is given by

\begin{equation}
\begin{split}
\eta_{\mathrm{t}} &=\frac{\gamma_{1}\left|a_{1}\right|^{2}+\gamma_{2}\left|a_{2}\right|^{2}}{\gamma_{0}|a|^{2}+\gamma_{1}\left|a_{1}\right|^{2}+\gamma_{2}\left|a_{2}\right|^{2}} \\
&=\frac{\gamma_{1}\left|\frac{K_{1}}{\sqrt{K_{1}^{2}+K_{2}^{2}}}\right|^{2}|a|^{2}+\gamma_{2}\left|\frac{K_{2}}{\sqrt{K_{1}^{2}+K_{2}^{2}}}\right|^{2}|a|^{2}}{\gamma_{0}|a|^{2}+\gamma_{1}\left|\frac{K_{1}}{\sqrt{K_{1}^{2}+K_{2}^{2}}}\right|^{2}|a|^{2}+\gamma_{2}\left|\frac{K_{2}}{\sqrt{K_{1}^{2}+K_{2}^{2}}}\right|^{2}|a|^{2}} \\
&=\frac{\gamma_{\mathrm{r}}}{\gamma_{\mathrm{r}}+\gamma_{0}}.
\end{split}
\end{equation}

Therefore, the total efficiency in the strong coupling region is {\em independent} of the couplings with the transmitter, as expected.

Analogously, for the individual receiver \#1 in the strong coupling $K_{1}>K_{1, e}$, the efficiency is written as

\begin{equation}
\begin{split}
\eta_{1} &=\frac{\gamma_{1}\left|a_{1}\right|^{2}}{\gamma_{0}|a|^{2}+\gamma_{1}\left|a_{1}\right|^{2}+\gamma_{2}\left|a_{2}\right|^{2}} \\
&=\frac{\gamma_{1}\left|\frac{K_{1}}{\sqrt{K_{1}^{2}+K_{2}^{2}}}\right|^{2}|a|^{2}}{\gamma_{0}|a|^{2}+\gamma_{1}\left|\frac{K_{1}}{\sqrt{K_{1}^{2}+K_{2}^{2}}}\right|^{2}|a|^{2}+\gamma_{2}\left|\frac{K_{2}}{\sqrt{K_{1}^{2}+K_{2}^{2}}}\right|^{2}|a|^{2}} \\
&=\frac{\gamma_{\mathrm{r}}}{\gamma_{\mathrm{r}}+\gamma_{0}} \frac{K_{1}^{2}}{K_{1}^{2}+K_{2}^{2}}.
\end{split}
\end{equation}

In the strong coupling regime, receiver \#1's efficiency depends on both coupling factors with the transmitter. However, depending on the value of $K_{2}$, the efficiency can remain approximately constant for $K_{2}<\gamma_{\mathrm{r}}$. Since in the strong coupling regime, the value of $K_{1}$ is much larger than $K_{2}$, there is no significant variation in efficiency. A similar analysis holds for receiver \#2 as well.

\section{Nonlinear Gain and Stability Analysis}
\label{App: B}
The gain element in the real circuit has a saturable nonlinearity. Its gain value saturates as the mode amplitude $|a|$ increases, resulting in stable oscillations. Among the few resonant modes available, the mode requiring the lowest gain will dominate, reaching its steady state and saturating the gain, which prevents other modes from accessing the necessary gain to achieve other steady-state oscillations. In this section, we provide an analysis of stability using the Lyapunov exponent behavior and demonstrate that the system settles into a steady state within a few cycles. The EPD condition we consider here is associated with the saturated gain value.

The saturated mode amplitudes in the transmitter $(\tilde{a})$ and receivers ($\tilde{a}_{1}$ and $\tilde{a}_{2}$) are denoted by a tilde and are associated with a steady state oscillation at a real-valued angular eigenfrequency $\omega$. If the system is slightly perturbed, we assume that the deviation from the steady state regime is denoted by $\rho \propto e^{\lambda t}$ for the transmitter and  $\rho_{1,2,} \propto e^{\lambda t}$ for the receivers, where $\lambda$ is the Lyapunov exponent. Our goal is to determine whether these perturbations quickly vanish over time, thereby assessing the stability of the system within a small neighborhood of the saturated regime. Therefore, for the two-receiver system shown in Fig. \ref{fig:2}, the signals are described by 
\begin{gather}
a(t) = \left(\tilde{a}+\rho(t)\right)e^{i\omega t}, \\
a_{1}(t) = \left(\tilde{a}_1 + \rho_1(t)\right)e^{i\omega t}, \\
a_{2}(t) = \left(\tilde{a}_2 + \rho_2(t)\right)e^{i\omega t}.   
\end{gather}
We assume that the saturable nonlinear gain is 
\begin{equation}
g(|a / \tilde{a}|)=-\gamma_{0}+\frac{2(\gamma+\gamma_{0})}{1+|a / \tilde{a}|^{2}}.
\end{equation} 
When the value of the mode amplitude in the transmitter $(a)$ reaches the saturation value $(\tilde{a})$, the saturated gain becomes $g=\gamma$ where
\begin{equation}
\gamma = \begin{cases} 
\left(K_1^2 + K_2^2 \right)/\gamma_{\mathrm{r}} & \sqrt{K_1^2 + K_2^2} < \gamma_{\mathrm{r}} \\ 
\gamma_{\mathrm{r}} & \sqrt{K_1^2 + K_2^2} \geq \gamma_{\mathrm{r}}
\end{cases},
\end{equation} 
in the weak and strong coupling regimes, respectively. 

Note that these values of saturated gain are compatible with having always one real solution as described after Eq. (\ref{eq:epd}), and this is important because a saturated regime of oscillation is associated with a real-valued eigenfrequency. Additionally, we recall that the EPD occurs when the gain satisfies $\gamma = \gamma_{\mathrm{r}}=\sqrt{K_1^2 + K_2^2}$.  If $\gamma_{\mathrm{r}}$ were used instead of $\gamma$, the system would become unstable in the weak coupling regime, as shown in Fig. \ref{fig:3}, due to a negative imaginary part of the eigenfrequency, leading to exponentially growing oscillations. Based on Eq. (\ref{eq:ODE_system}), coupled-mode theory for this system, assuming $K_{12}=0$, and considering the nonlinear gain model, is rewritten as

\begin{equation}
\frac{d}{d t}\left[\begin{array}{c}
a \\
a_1 \\
a_2
\end{array}\right]=\left[\begin{array}{ccc}
i \omega_0+g\left(|a/\tilde{a}|\right) & -i K_1 & -i K_2 \\
-i K_1 & i \omega_1-\gamma_1 & 0 \\
-i K_2 & 0 & i \omega_2-\gamma_2
\end{array}\right]\left[\begin{array}{c}
a \\
a_1 \\
a_2
\end{array}\right].
\label{eq:ODE_system_NL}
\end{equation}

To solve it, we simplify the gain model around the steady-state response using Taylor expansion for the small perturbation $\rho$. Neglecting the quadratic terms of $\rho$, this procedure leads to

\begin{equation}
\begin{split}
g(|a / \tilde{a}|) & = -\gamma_{0}+\frac{2\left(\gamma+\gamma_{0}\right)}{1+(\tilde{a}+\rho)(\tilde{a}+\rho)^{*} /|\tilde{a}|^{2}} \\
& \approx -\gamma_{0}+\frac{\gamma+\gamma_{0}}{1+\left(\tilde{a}\rho^*+\tilde{a}^*\rho\right) /\left(2|\tilde{a}|^2\right)} \\
& \approx -\gamma_{0}+\left(\gamma+\gamma_{0}\right)\left(1-\frac{\tilde{a} \rho^{*}+\widetilde{a^{*}} \rho}{2|\tilde{a}|^{2}}\right).    
\end{split}
\label{eq:linearized_gain}
\end{equation}

The differential equation for the transmitter is written as

\begin{equation}
    \frac{d}{d t}a=\left[i\omega_{0}+g(|a / \tilde{a}|)\right]a -i K_{1}a_1-i K_{2}a_2.    
\end{equation}

which leads to

\begin{equation}
\begin{split}
    \frac{d \rho}{d t} +i\omega(\tilde{a} + \rho)=\left[i\omega_{0}+g(|a / \tilde{a}|)\right]\left(\tilde{a} + \rho \right) \\
    -i K_{1}\left(\tilde{a}_1 + \rho_1 \right)-i K_{2}\left(\tilde{a}_2 + \rho_2 \right).
\end{split}
\end{equation}

Substituting for the simplified gain derived in Eq. (\ref{eq:linearized_gain}) leads to

\begin{equation}
\begin{split}
    \frac{d \rho}{d t} = \left[i\left(\omega_{0} - \omega \right) -\gamma_{0}+\left(\gamma+\gamma_{0}\right) \right]\left(\tilde{a} + \rho \right) \\
    -\left[\left(\gamma+\gamma_{0}\right) \left(\frac{\tilde{a} \rho^{*}+\widetilde{a^{*}} \rho}{2|\tilde{a}|^{2}}\right) \right]\left(\tilde{a} + \rho \right) \\
    -i K_{1}\left(\tilde{a}_1 + \rho_1 \right)-i K_{2}\left(\tilde{a}_2 + \rho_2 \right).
\end{split}
\end{equation}

Using the coupled-mode theory equation for the transmitter, as given in Eq. \eqref{eq:transmitter}, and omitting the quadratic terms in $\rho$, leads to

\begin{equation}
\begin{split}
    \frac{d \rho}{d t} = \left[i\left(\omega_{0} - \omega \right) -\gamma_{0}+\left(\gamma+\gamma_{0}\right) \right]\rho \\
    -\left[\left(\gamma+\gamma_{0}\right) \left(\frac{\tilde{a} \rho^{*}+\widetilde{a^{*}} \rho}{2|\tilde{a}|^{2}}\right) \right]\tilde{a} \\
    -i K_{1} \rho_1 -i K_{2} \rho_2.
\end{split}
\end{equation}

In a similar manner to the approach used for the first equation in the system of differential equations in Eq. \eqref{eq:ODE_system_NL}, we can derive the three linearized differential equations for the perturbations, given by

\begin{gather}
\frac{d}{d t} \rho=B_{1} \rho+B_{2} \rho^{*}+C_{1} \rho_{1}+C_{2} \rho_{2}, \\
\frac{d}{d t} \rho_{1}=D_{1} \rho_{1}+C_{1} \rho, \\
\frac{d}{d t} \rho_{2}=D_{2} \rho_{2}+C_{2} \rho, 
\end{gather}
where

\begin{gather}
B_{1}=i\left(\omega_{0}-\omega\right) - \gamma_0 +\frac{1}{2}\left(\gamma+\gamma_{0}\right), \\
B_{2}=-\frac{1}{2}\left(\gamma+\gamma_{0}\right) \frac{\tilde{a}^{2}}{|\tilde{a}|^{2}}, \\
C_{1}=-i K_{1}, \\
C_{2}=-i K_{2}, \\
D_{1}=i\left(\omega_1-\omega\right)-\gamma_{\mathrm{r}}, \\
D_{2}=i\left(\omega_2-\omega\right)-\gamma_{\mathrm{r}}.
\end{gather}

Here, we assume that the resonance frequencies of the uncoupled resonators are matched, i.e., $\omega_0 = \omega_1 = \omega_2$. Additionally, as mentioned earlier, $\omega$ refers to the saturated regime oscillation frequency, meaning that $\omega$ corresponds to the real-valued eigenfrequency derived in Eq. (\ref{eq:eigenfrequency}) with gain as in Eq. \eqref{eq:gsaturated}. This means that 
\begin{equation}
\omega = \omega_0 - \mathrm{Re}\left(\sqrt{K_1^2 + K_2^2 - \gamma_{\mathrm{r}}^2}\right), 
\label{eq:satfreq}
\end{equation}
which corresponds to the measured oscillation frequency of the system shown in Fig. \ref{fig:7}(b). Furthermore, we assume that the perturbations have the time dependency as \cite{zhou2016pt}

\begin{gather}
    \rho  =u e^{\lambda t}+v^{*} e^{\lambda^{*} t}, \\
    \rho_{1} =u_{1} e^{\lambda t}+v_{1}^{*} e^{\lambda^{*} t}, \\
    \rho_{2}  =u_{2} e^{\lambda t}+v_{2}^{*} e^{\lambda^{*} t}.
\end{gather}

Now, the linear system in matrix form, derived from the above differential equations, is

\begin{equation}
\left[\begin{array}{cccccc}
B_{1} & B_{2} & C_{1} & 0 & C_{2} & 0  \\
B_{2}{ }^{*} & B_{1}{ }^{*} & 0 & C_{1}^{*} & 0 & C_{2}^{*} \\
C_{1} & 0 & D_{1} & 0 & 0 & 0 \\
0 & C_{1}^{*} & 0 & D_{1}^{*} & 0 & 0 \\
C_{2} & 0 & 0 & 0 & D_{2} & 0 \\
0 & C_{2}^{*} & 0 & 0 & 0 & D_{2}^{*}
\end{array}\right]\left[\begin{array}{c}
u \\
v \\
u_{1} \\
v_{1} \\
u_{2} \\
v_{2}
\end{array}\right]=\lambda\left[\begin{array}{c}
u \\
v \\
u_{1} \\
v_{1} \\
u_{2} \\
v_{2}
\end{array}\right].
\label{eq:Diff_Matrix}
\end{equation}

\begin{figure}
\centering	
\includegraphics[width=1\columnwidth]{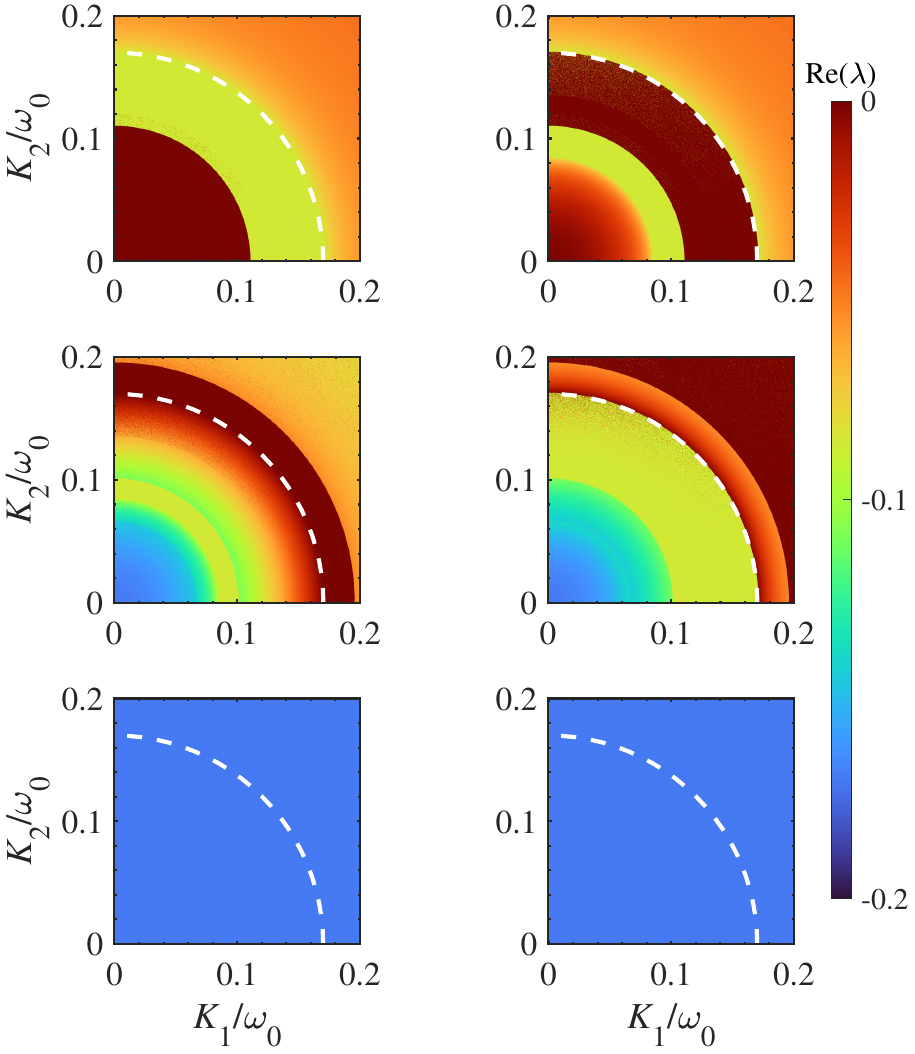}
\caption{Real part of the Lyapunov exponents (six solutions of Eq. \eqref{eq:Diff_Matrix}) to confirm the steady-state solution, by varying both $K_{1}$ and $K_{2}$, assuming $\omega_0 = \omega_1=\omega_2$. For all coupling cases, the real part is always negative, which implies a decaying perturbation, always converging to the saturated regime denoted by the tilde symbol. In other, words, the system remains stable for any set of coupling coefficients considered. The circular white dashed line represents the EPD condition $K_{2}^{2}+K_{1}^{2}=\gamma_{\mathrm{r}}^{2}$ with the values of $\gamma_{\mathrm{r}} / \omega_{0}=$ 0.17 and $\gamma_{0} / \omega_{0}=0.001$, separating the weak and strong coupling regimes.}
\label{fig:8}
\end{figure}

We now calculate the six eigenvalues $(\lambda)$  that are shown in  Fig. \ref{fig:8} varying the two coupling coefficients, i.e., spanning weak and strong coupling regimes. For each pair of chosen values of $K_1$ and $K_2$, we assume that  $\omega$ is provided by Eq. (\ref{eq:satfreq}).
The calculated Lyapunov exponents $\lambda$ are all decaying types $\left(\mathrm{Re}(\lambda)<0 \right)$, indicating that the steady-state solution is stable for all transfer distances for both receivers, i.e., for any weak or strong coupling regime.

\section{Power Transfer Measurement Setup}
\label{App: C}
The measurement setup includes a 68-turn Litz wire transmitter coil with a $30\:\mathrm{mm}$ diameter, quality factor of $Q_{\text{Tx}}=720$, and intrinsic loss of $0.55\:\Omega$. Two 35-turn receiver coils were also fabricated with Litz wire with $10\:\mathrm{mm}$ diameter and quality factor $Q_{\text{Rx}}=680$ and intrinsic loss of $0.078\:\Omega$. The small value for intrinsic losses in the coils validates the initial assumption that the intrinsic losses of inductors in the receivers are negligible compared to the losses in the loads $G_1$ and $G_2$. The value $G_{0}=0.002\:\mathrm{mS}$, is the parallel conversion of series loss in the transmitter coil, at $\omega_{0}$ and assumed constant for simplicity.

\begin{figure}
\centering	
\includegraphics[width=1\columnwidth]{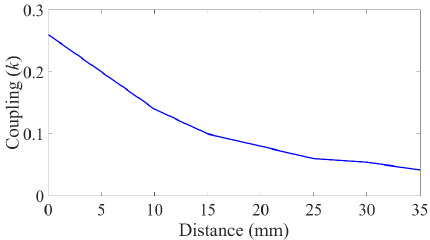}
\caption{Mutual inductance coupling factor by varying the distance between the transmitter and a receiver.}
\label{fig:9}
\end{figure}

\begin{figure}
\centering
\includegraphics[width=1\columnwidth]{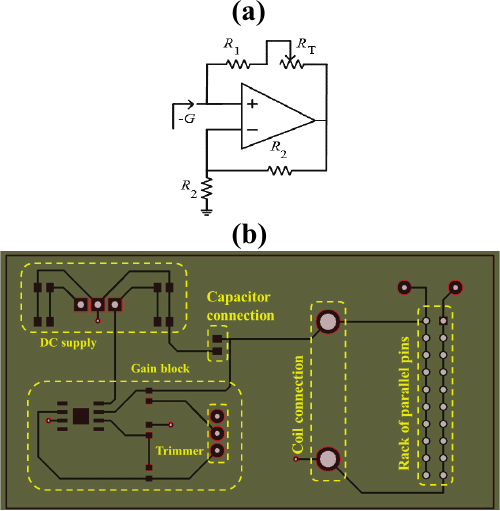}
\caption{(a) Op-Amp configuration to realize the gain $-G_{\text {gain }}=$ $-1 /\left(R_{1}+R_{\mathrm{T}}\right)$, where the trimmer $R_{\mathrm{T}}$ is tuned to get the gain value for the system to have an EPD. (b) Printed circuit board layout of the transmitter's circuit where the traces are black, the ground plane underneath is green, and the connecting vias are brown. The blocks in yellow are the DC supply, Gain block, Coil connection, and rack of parallel pins for capacitance adjustments.}
\label{fig:10}
\end{figure}

The coupling factor change based on the distance between the transmitter and one receiver coil is shown in Fig. \ref{fig:9}. The coupling levels between coils were calculated using the COMSOL Multiphysics for magnetic field simulation of the measuring setup. Transmitter and receivers' inductors coils are in parallel with $330\:\mathrm{pF}$ and $3.3\:\mathrm{nF}$ capacitors, respectively. In addition, a rack of parallel pins is installed on the transmitter board to add discrete capacitors to tune the resonant frequency of the transmitter. The transmitter part is terminated with a gain realized with an Op-Amp-based circuit shown in Fig. \ref{fig:10}(a).

To realize and tune the negative gain amount, we used an Op-Amp (Analog Devices, model ADA4817), where the negative value is tuned with a trimmer $R_{\mathrm{T}}$ (Bourns, model 3252W-1-103LF). This block provides negative admittance $-G_{\text {gain }}=-1 /\left(R_{1}+R_{\mathrm{T}}\right)$, where $R_{1}$ is the mount fixed resistance. The $R_{T}$ is a trimmer resistance to tune the total feedback resistance $\left(R_{1}+R_{\mathrm{T}}\right)$ in the circuit and bring the negative conductance close to the EPD value based on the required gain for EPD in the theoretical calculation. After trimming the value of $R_{\mathrm{T}}$ for the required gain, it has been kept fixed in all the experiments. Figure. \ref{fig:10}(b) shows the printed circuit board (PCB) of the assembled circuit, where each block is shown in yellow boxes. All the ground nodes are connected using the bottom green ground layer.

\section{EPD Condition for N-Receivers}
The concepts provided in this paper can be easily generalized to the $N$ number of receivers while maintaining a constant power transfer efficiency for each receiver in their strong coupling region since even with $N$ number of non-identical receivers, the EPD point still exists. The eigenfrequencies for a $N+1$ resonator system, assuming $g=\gamma_{\mathrm{r}}$, is given by

\begin{figure}[b!]
\centering	
\includegraphics[width=1\columnwidth]{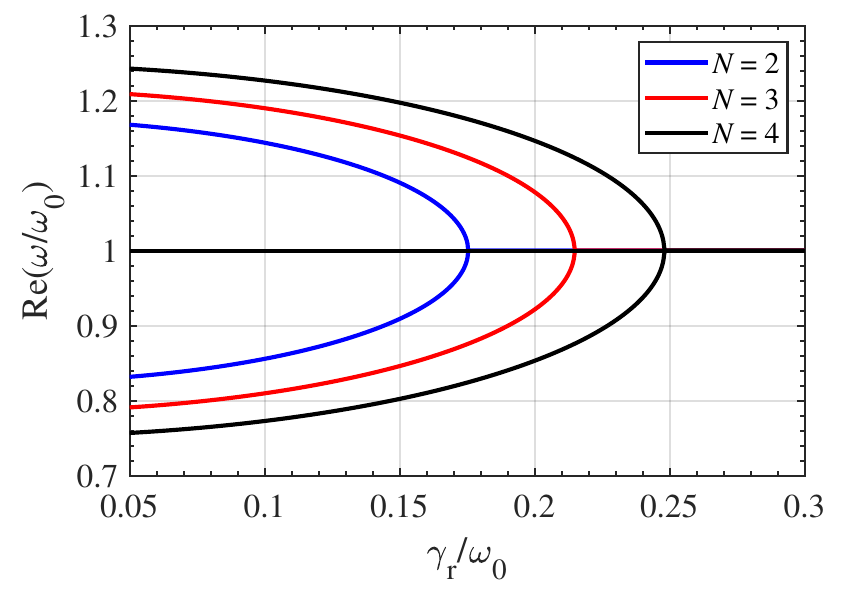}
\caption{Real part of the angular eigenfrequencies when there are $N$ receivers in the system. An EPD can still be found while $K_{1}/\omega_{0}= K_{2} / \omega_{0} = \cdots=K_{N} / \omega_{0}=0.124$. Here, $N$ represents the number of receivers.}
\label{fig:11}
\end{figure}

\begin{equation}
   \omega = 
     \begin{cases}
        \omega_{0}+i \gamma_{\mathrm{r}}\\
        \omega_{0}-\sqrt{K_{1}^{2}+K_{2}^{2}+\cdots+K_{N}^{2}-\gamma_{\mathrm{r}}^{2}} \\
        \omega_{0}+\sqrt{K_{1}^{2}+K_{2}^{2}+\cdots+K_{N}^{2}-\gamma_{\mathrm{r}}^{2}}
     \end{cases}
     \label{eq:N_receivers},
\end{equation}
with $\gamma_{\text{epd}}=\sqrt{K_{1}^{2}+K_{2}^{2}+\cdots+K_{N}^{2}}$. \textcolor{black}{Note that the multiplicity of first eigenfrequency is $(N-1)$}. Although adding more receivers shortens the "high efficiency" region, as shown in Fig. \ref{fig:11}, still, in the strong coupling range, the system sensitivity to disturbances that can change the coupling factor between transmitter and receivers will be reduced.

% \bibliography{Reference}% Produces the bibliography via BibTeX.

\begin{thebibliography}{43}%
\makeatletter
\providecommand \@ifxundefined [1]{%
 \@ifx{#1\undefined}
}%
\providecommand \@ifnum [1]{%
 \ifnum #1\expandafter \@firstoftwo
 \else \expandafter \@secondoftwo
 \fi
}%
\providecommand \@ifx [1]{%
 \ifx #1\expandafter \@firstoftwo
 \else \expandafter \@secondoftwo
 \fi
}%
\providecommand \natexlab [1]{#1}%
\providecommand \enquote  [1]{``#1''}%
\providecommand \bibnamefont  [1]{#1}%
\providecommand \bibfnamefont [1]{#1}%
\providecommand \citenamefont [1]{#1}%
\providecommand \href@noop [0]{\@secondoftwo}%
\providecommand \href [0]{\begingroup \@sanitize@url \@href}%
\providecommand \@href[1]{\@@startlink{#1}\@@href}%
\providecommand \@@href[1]{\endgroup#1\@@endlink}%
\providecommand \@sanitize@url [0]{\catcode `\\12\catcode `\$12\catcode
  `\&12\catcode `\#12\catcode `\^12\catcode `\_12\catcode `\%12\relax}%
\providecommand \@@startlink[1]{}%
\providecommand \@@endlink[0]{}%
\providecommand \url  [0]{\begingroup\@sanitize@url \@url }%
\providecommand \@url [1]{\endgroup\@href {#1}{\urlprefix }}%
\providecommand \urlprefix  [0]{URL }%
\providecommand \Eprint [0]{\href }%
\providecommand \doibase [0]{https://doi.org/}%
\providecommand \selectlanguage [0]{\@gobble}%
\providecommand \bibinfo  [0]{\@secondoftwo}%
\providecommand \bibfield  [0]{\@secondoftwo}%
\providecommand \translation [1]{[#1]}%
\providecommand \BibitemOpen [0]{}%
\providecommand \bibitemStop [0]{}%
\providecommand \bibitemNoStop [0]{.\EOS\space}%
\providecommand \EOS [0]{\spacefactor3000\relax}%
\providecommand \BibitemShut  [1]{\csname bibitem#1\endcsname}%
\let\auto@bib@innerbib\@empty
%</preamble>
\bibitem [{\citenamefont {Sample}\ \emph {et~al.}(2010)\citenamefont {Sample},
  \citenamefont {Meyer},\ and\ \citenamefont {Smith}}]{sample2010analysis}%
  \BibitemOpen
  \bibfield  {author} {\bibinfo {author} {\bibfnamefont {A.~P.}\ \bibnamefont
  {Sample}}, \bibinfo {author} {\bibfnamefont {D.~T.}\ \bibnamefont {Meyer}},\
  and\ \bibinfo {author} {\bibfnamefont {J.~R.}\ \bibnamefont {Smith}},\
  }\bibfield  {title} {\bibinfo {title} {Analysis, experimental results, and
  range adaptation of magnetically coupled resonators for wireless power
  transfer},\ }\href@noop {} {\bibfield  {journal} {\bibinfo  {journal} {IEEE
  Transactions on industrial electronics}\ }\textbf {\bibinfo {volume} {58}},\
  \bibinfo {pages} {544} (\bibinfo {year} {2010})}\BibitemShut {NoStop}%
\bibitem [{\citenamefont {Song}\ \emph {et~al.}(2021)\citenamefont {Song},
  \citenamefont {Jayathurathnage}, \citenamefont {Zanganeh}, \citenamefont
  {Krasikova}, \citenamefont {Smirnov}, \citenamefont {Belov}, \citenamefont
  {Kapitanova}, \citenamefont {Simovski}, \citenamefont {Tretyakov},\ and\
  \citenamefont {Krasnok}}]{song2021wireless}%
  \BibitemOpen
  \bibfield  {author} {\bibinfo {author} {\bibfnamefont {M.}~\bibnamefont
  {Song}}, \bibinfo {author} {\bibfnamefont {P.}~\bibnamefont
  {Jayathurathnage}}, \bibinfo {author} {\bibfnamefont {E.}~\bibnamefont
  {Zanganeh}}, \bibinfo {author} {\bibfnamefont {M.}~\bibnamefont {Krasikova}},
  \bibinfo {author} {\bibfnamefont {P.}~\bibnamefont {Smirnov}}, \bibinfo
  {author} {\bibfnamefont {P.}~\bibnamefont {Belov}}, \bibinfo {author}
  {\bibfnamefont {P.}~\bibnamefont {Kapitanova}}, \bibinfo {author}
  {\bibfnamefont {C.}~\bibnamefont {Simovski}}, \bibinfo {author}
  {\bibfnamefont {S.}~\bibnamefont {Tretyakov}},\ and\ \bibinfo {author}
  {\bibfnamefont {A.}~\bibnamefont {Krasnok}},\ }\bibfield  {title} {\bibinfo
  {title} {Wireless power transfer based on novel physical concepts},\
  }\href@noop {} {\bibfield  {journal} {\bibinfo  {journal} {Nature
  Electronics}\ }\textbf {\bibinfo {volume} {4}},\ \bibinfo {pages} {707}
  (\bibinfo {year} {2021})}\BibitemShut {NoStop}%
\bibitem [{\citenamefont {Hui}\ \emph {et~al.}(2023)\citenamefont {Hui},
  \citenamefont {Yang},\ and\ \citenamefont {Zhang}}]{hui2023wireless}%
  \BibitemOpen
  \bibfield  {author} {\bibinfo {author} {\bibfnamefont {S.-Y.~R.}\
  \bibnamefont {Hui}}, \bibinfo {author} {\bibfnamefont {Y.}~\bibnamefont
  {Yang}},\ and\ \bibinfo {author} {\bibfnamefont {C.}~\bibnamefont {Zhang}},\
  }\bibfield  {title} {\bibinfo {title} {Wireless power transfer: A paradigm
  shift for the next generation},\ }\href@noop {} {\bibfield  {journal}
  {\bibinfo  {journal} {IEEE Journal of Emerging and Selected Topics in Power
  Electronics}\ }\textbf {\bibinfo {volume} {11}},\ \bibinfo {pages} {2412}
  (\bibinfo {year} {2023})}\BibitemShut {NoStop}%
\bibitem [{\citenamefont {Kurs}\ \emph {et~al.}(2007)\citenamefont {Kurs},
  \citenamefont {Karalis}, \citenamefont {Moffatt}, \citenamefont
  {Joannopoulos}, \citenamefont {Fisher},\ and\ \citenamefont
  {Soljacic}}]{kurs2007wireless}%
  \BibitemOpen
  \bibfield  {author} {\bibinfo {author} {\bibfnamefont {A.}~\bibnamefont
  {Kurs}}, \bibinfo {author} {\bibfnamefont {A.}~\bibnamefont {Karalis}},
  \bibinfo {author} {\bibfnamefont {R.}~\bibnamefont {Moffatt}}, \bibinfo
  {author} {\bibfnamefont {J.~D.}\ \bibnamefont {Joannopoulos}}, \bibinfo
  {author} {\bibfnamefont {P.}~\bibnamefont {Fisher}},\ and\ \bibinfo {author}
  {\bibfnamefont {M.}~\bibnamefont {Soljacic}},\ }\bibfield  {title} {\bibinfo
  {title} {Wireless power transfer via strongly coupled magnetic resonances},\
  }\href@noop {} {\bibfield  {journal} {\bibinfo  {journal} {science}\ }\textbf
  {\bibinfo {volume} {317}},\ \bibinfo {pages} {83} (\bibinfo {year}
  {2007})}\BibitemShut {NoStop}%
\bibitem [{\citenamefont {Liu}\ \emph {et~al.}(2023)\citenamefont {Liu},
  \citenamefont {Chau}, \citenamefont {Tian}, \citenamefont {Wang},\ and\
  \citenamefont {Hua}}]{liu2023smart}%
  \BibitemOpen
  \bibfield  {author} {\bibinfo {author} {\bibfnamefont {W.}~\bibnamefont
  {Liu}}, \bibinfo {author} {\bibfnamefont {K.}~\bibnamefont {Chau}}, \bibinfo
  {author} {\bibfnamefont {X.}~\bibnamefont {Tian}}, \bibinfo {author}
  {\bibfnamefont {H.}~\bibnamefont {Wang}},\ and\ \bibinfo {author}
  {\bibfnamefont {Z.}~\bibnamefont {Hua}},\ }\bibfield  {title} {\bibinfo
  {title} {Smart wireless power transfer—opportunities and challenges},\
  }\href@noop {} {\bibfield  {journal} {\bibinfo  {journal} {Renewable and
  Sustainable Energy Reviews}\ }\textbf {\bibinfo {volume} {180}},\ \bibinfo
  {pages} {113298} (\bibinfo {year} {2023})}\BibitemShut {NoStop}%
\bibitem [{\citenamefont {Hajiaghajani}\ \emph {et~al.}(2019)\citenamefont
  {Hajiaghajani}, \citenamefont {Kim}, \citenamefont {Abdolali},\ and\
  \citenamefont {Ahn}}]{hajiaghajani2019patterned}%
  \BibitemOpen
  \bibfield  {author} {\bibinfo {author} {\bibfnamefont {A.}~\bibnamefont
  {Hajiaghajani}}, \bibinfo {author} {\bibfnamefont {D.}~\bibnamefont {Kim}},
  \bibinfo {author} {\bibfnamefont {A.}~\bibnamefont {Abdolali}},\ and\
  \bibinfo {author} {\bibfnamefont {S.}~\bibnamefont {Ahn}},\ }\bibfield
  {title} {\bibinfo {title} {Patterned magnetic fields for remote steering and
  wireless powering to a swimming microrobot},\ }\href@noop {} {\bibfield
  {journal} {\bibinfo  {journal} {IEEE/ASME Transactions on Mechatronics}\
  }\textbf {\bibinfo {volume} {25}},\ \bibinfo {pages} {207} (\bibinfo {year}
  {2019})}\BibitemShut {NoStop}%
\bibitem [{\citenamefont {Sakhdari}\ \emph {et~al.}(2020)\citenamefont
  {Sakhdari}, \citenamefont {Hajizadegan},\ and\ \citenamefont
  {Chen}}]{sakhdari2020robust}%
  \BibitemOpen
  \bibfield  {author} {\bibinfo {author} {\bibfnamefont {M.}~\bibnamefont
  {Sakhdari}}, \bibinfo {author} {\bibfnamefont {M.}~\bibnamefont
  {Hajizadegan}},\ and\ \bibinfo {author} {\bibfnamefont {P.-Y.}\ \bibnamefont
  {Chen}},\ }\bibfield  {title} {\bibinfo {title} {Robust extended-range
  wireless power transfer using a higher-order {PT}-symmetric platform},\
  }\href@noop {} {\bibfield  {journal} {\bibinfo  {journal} {Physical Review
  Research}\ }\textbf {\bibinfo {volume} {2}},\ \bibinfo {pages} {013152}
  (\bibinfo {year} {2020})}\BibitemShut {NoStop}%
\bibitem [{\citenamefont {Shu}\ \emph {et~al.}(2018)\citenamefont {Shu},
  \citenamefont {Xiao},\ and\ \citenamefont {Zhang}}]{shu2018wireless}%
  \BibitemOpen
  \bibfield  {author} {\bibinfo {author} {\bibfnamefont {X.}~\bibnamefont
  {Shu}}, \bibinfo {author} {\bibfnamefont {W.}~\bibnamefont {Xiao}},\ and\
  \bibinfo {author} {\bibfnamefont {B.}~\bibnamefont {Zhang}},\ }\bibfield
  {title} {\bibinfo {title} {Wireless power supply for small household
  appliances using energy model},\ }\href@noop {} {\bibfield  {journal}
  {\bibinfo  {journal} {IEEE Access}\ }\textbf {\bibinfo {volume} {6}},\
  \bibinfo {pages} {69592} (\bibinfo {year} {2018})}\BibitemShut {NoStop}%
\bibitem [{\citenamefont {Mohseni}\ \emph {et~al.}(2021)\citenamefont
  {Mohseni}, \citenamefont {Marsh}, \citenamefont {Capolino}, \citenamefont
  {Chiao},\ and\ \citenamefont {Cao}}]{mohseni2021design}%
  \BibitemOpen
  \bibfield  {author} {\bibinfo {author} {\bibfnamefont {F.}~\bibnamefont
  {Mohseni}}, \bibinfo {author} {\bibfnamefont {P.}~\bibnamefont {Marsh}},
  \bibinfo {author} {\bibfnamefont {F.}~\bibnamefont {Capolino}}, \bibinfo
  {author} {\bibfnamefont {J.-C.}\ \bibnamefont {Chiao}},\ and\ \bibinfo
  {author} {\bibfnamefont {H.}~\bibnamefont {Cao}},\ }\bibfield  {title}
  {\bibinfo {title} {Design of a wireless power and data transfer system for ph
  sensing inside a small tube},\ }in\ \href@noop {} {\emph {\bibinfo
  {booktitle} {2021 IEEE Wireless Power Transfer Conference (WPTC)}}}\
  (\bibinfo {organization} {IEEE},\ \bibinfo {year} {2021})\ pp.\ \bibinfo
  {pages} {1--4}\BibitemShut {NoStop}%
\bibitem [{\citenamefont {Karalis}\ \emph {et~al.}(2008)\citenamefont
  {Karalis}, \citenamefont {Joannopoulos},\ and\ \citenamefont
  {Solja{\v{c}}i{\'c}}}]{karalis2008efficient}%
  \BibitemOpen
  \bibfield  {author} {\bibinfo {author} {\bibfnamefont {A.}~\bibnamefont
  {Karalis}}, \bibinfo {author} {\bibfnamefont {J.~D.}\ \bibnamefont
  {Joannopoulos}},\ and\ \bibinfo {author} {\bibfnamefont {M.}~\bibnamefont
  {Solja{\v{c}}i{\'c}}},\ }\bibfield  {title} {\bibinfo {title} {Efficient
  wireless non-radiative mid-range energy transfer},\ }\href@noop {} {\bibfield
   {journal} {\bibinfo  {journal} {Annals of physics}\ }\textbf {\bibinfo
  {volume} {323}},\ \bibinfo {pages} {34} (\bibinfo {year} {2008})}\BibitemShut
  {NoStop}%
\bibitem [{\citenamefont {Kiani}\ \emph {et~al.}(2011)\citenamefont {Kiani},
  \citenamefont {Jow},\ and\ \citenamefont {Ghovanloo}}]{kiani2011design}%
  \BibitemOpen
  \bibfield  {author} {\bibinfo {author} {\bibfnamefont {M.}~\bibnamefont
  {Kiani}}, \bibinfo {author} {\bibfnamefont {U.-M.}\ \bibnamefont {Jow}},\
  and\ \bibinfo {author} {\bibfnamefont {M.}~\bibnamefont {Ghovanloo}},\
  }\bibfield  {title} {\bibinfo {title} {Design and optimization of a 3-coil
  inductive link for efficient wireless power transmission},\ }\href@noop {}
  {\bibfield  {journal} {\bibinfo  {journal} {IEEE transactions on biomedical
  circuits and systems}\ }\textbf {\bibinfo {volume} {5}},\ \bibinfo {pages}
  {579} (\bibinfo {year} {2011})}\BibitemShut {NoStop}%
\bibitem [{\citenamefont {Ho}\ \emph {et~al.}(2013)\citenamefont {Ho},
  \citenamefont {Kim},\ and\ \citenamefont {Poon}}]{ho2013midfield}%
  \BibitemOpen
  \bibfield  {author} {\bibinfo {author} {\bibfnamefont {J.~S.}\ \bibnamefont
  {Ho}}, \bibinfo {author} {\bibfnamefont {S.}~\bibnamefont {Kim}},\ and\
  \bibinfo {author} {\bibfnamefont {A.~S.}\ \bibnamefont {Poon}},\ }\bibfield
  {title} {\bibinfo {title} {Midfield wireless powering for implantable
  systems},\ }\href@noop {} {\bibfield  {journal} {\bibinfo  {journal}
  {Proceedings of the IEEE}\ }\textbf {\bibinfo {volume} {101}},\ \bibinfo
  {pages} {1369} (\bibinfo {year} {2013})}\BibitemShut {NoStop}%
\bibitem [{\citenamefont {Schindler}\ \emph {et~al.}(2011)\citenamefont
  {Schindler}, \citenamefont {Li}, \citenamefont {Zheng}, \citenamefont
  {Ellis},\ and\ \citenamefont {Kottos}}]{schindler2011experimental}%
  \BibitemOpen
  \bibfield  {author} {\bibinfo {author} {\bibfnamefont {J.}~\bibnamefont
  {Schindler}}, \bibinfo {author} {\bibfnamefont {A.}~\bibnamefont {Li}},
  \bibinfo {author} {\bibfnamefont {M.~C.}\ \bibnamefont {Zheng}}, \bibinfo
  {author} {\bibfnamefont {F.~M.}\ \bibnamefont {Ellis}},\ and\ \bibinfo
  {author} {\bibfnamefont {T.}~\bibnamefont {Kottos}},\ }\bibfield  {title}
  {\bibinfo {title} {Experimental study of active lrc circuits with {PT}
  symmetries},\ }\href@noop {} {\bibfield  {journal} {\bibinfo  {journal}
  {Physical Review A}\ }\textbf {\bibinfo {volume} {84}},\ \bibinfo {pages}
  {040101} (\bibinfo {year} {2011})}\BibitemShut {NoStop}%
\bibitem [{\citenamefont {Heiss}(2012)}]{heiss2012physics}%
  \BibitemOpen
  \bibfield  {author} {\bibinfo {author} {\bibfnamefont {W.}~\bibnamefont
  {Heiss}},\ }\bibfield  {title} {\bibinfo {title} {The physics of exceptional
  points},\ }\href@noop {} {\bibfield  {journal} {\bibinfo  {journal} {Journal
  of Physics A: Mathematical and Theoretical}\ }\textbf {\bibinfo {volume}
  {45}},\ \bibinfo {pages} {444016} (\bibinfo {year} {2012})}\BibitemShut
  {NoStop}%
\bibitem [{\citenamefont {Assawaworrarit}\ \emph {et~al.}(2017)\citenamefont
  {Assawaworrarit}, \citenamefont {Yu},\ and\ \citenamefont
  {Fan}}]{assawaworrarit2017robust}%
  \BibitemOpen
  \bibfield  {author} {\bibinfo {author} {\bibfnamefont {S.}~\bibnamefont
  {Assawaworrarit}}, \bibinfo {author} {\bibfnamefont {X.}~\bibnamefont {Yu}},\
  and\ \bibinfo {author} {\bibfnamefont {S.}~\bibnamefont {Fan}},\ }\bibfield
  {title} {\bibinfo {title} {Robust wireless power transfer using a nonlinear
  parity--time-symmetric circuit},\ }\href@noop {} {\bibfield  {journal}
  {\bibinfo  {journal} {Nature}\ }\textbf {\bibinfo {volume} {546}},\ \bibinfo
  {pages} {387} (\bibinfo {year} {2017})}\BibitemShut {NoStop}%
\bibitem [{\citenamefont {Stehmann}\ \emph {et~al.}(2004)\citenamefont
  {Stehmann}, \citenamefont {Heiss},\ and\ \citenamefont
  {Scholtz}}]{stehmann2004observation}%
  \BibitemOpen
  \bibfield  {author} {\bibinfo {author} {\bibfnamefont {T.}~\bibnamefont
  {Stehmann}}, \bibinfo {author} {\bibfnamefont {W.}~\bibnamefont {Heiss}},\
  and\ \bibinfo {author} {\bibfnamefont {F.}~\bibnamefont {Scholtz}},\
  }\bibfield  {title} {\bibinfo {title} {Observation of exceptional points in
  electronic circuits},\ }\href@noop {} {\bibfield  {journal} {\bibinfo
  {journal} {Journal of Physics A: Mathematical and General}\ }\textbf
  {\bibinfo {volume} {37}},\ \bibinfo {pages} {7813} (\bibinfo {year}
  {2004})}\BibitemShut {NoStop}%
\bibitem [{\citenamefont {El-Ganainy}\ \emph {et~al.}(2007)\citenamefont
  {El-Ganainy}, \citenamefont {Makris}, \citenamefont {Christodoulides},\ and\
  \citenamefont {Musslimani}}]{el2007theory}%
  \BibitemOpen
  \bibfield  {author} {\bibinfo {author} {\bibfnamefont {R.}~\bibnamefont
  {El-Ganainy}}, \bibinfo {author} {\bibfnamefont {K.}~\bibnamefont {Makris}},
  \bibinfo {author} {\bibfnamefont {D.}~\bibnamefont {Christodoulides}},\ and\
  \bibinfo {author} {\bibfnamefont {Z.~H.}\ \bibnamefont {Musslimani}},\
  }\bibfield  {title} {\bibinfo {title} {Theory of coupled optical
  {PT}-symmetric structures},\ }\href@noop {} {\bibfield  {journal} {\bibinfo
  {journal} {Optics letters}\ }\textbf {\bibinfo {volume} {32}},\ \bibinfo
  {pages} {2632} (\bibinfo {year} {2007})}\BibitemShut {NoStop}%
\bibitem [{\citenamefont {Hodaei}\ \emph {et~al.}(2014)\citenamefont {Hodaei},
  \citenamefont {Miri}, \citenamefont {Heinrich}, \citenamefont
  {Christodoulides},\ and\ \citenamefont {Khajavikhan}}]{hodaei2014parity}%
  \BibitemOpen
  \bibfield  {author} {\bibinfo {author} {\bibfnamefont {H.}~\bibnamefont
  {Hodaei}}, \bibinfo {author} {\bibfnamefont {M.-A.}\ \bibnamefont {Miri}},
  \bibinfo {author} {\bibfnamefont {M.}~\bibnamefont {Heinrich}}, \bibinfo
  {author} {\bibfnamefont {D.~N.}\ \bibnamefont {Christodoulides}},\ and\
  \bibinfo {author} {\bibfnamefont {M.}~\bibnamefont {Khajavikhan}},\
  }\bibfield  {title} {\bibinfo {title} {Parity-time--symmetric microring
  lasers},\ }\href@noop {} {\bibfield  {journal} {\bibinfo  {journal}
  {Science}\ }\textbf {\bibinfo {volume} {346}},\ \bibinfo {pages} {975}
  (\bibinfo {year} {2014})}\BibitemShut {NoStop}%
\bibitem [{\citenamefont {Feng}\ \emph {et~al.}(2014)\citenamefont {Feng},
  \citenamefont {Wong}, \citenamefont {Ma}, \citenamefont {Wang},\ and\
  \citenamefont {Zhang}}]{feng2014single}%
  \BibitemOpen
  \bibfield  {author} {\bibinfo {author} {\bibfnamefont {L.}~\bibnamefont
  {Feng}}, \bibinfo {author} {\bibfnamefont {Z.~J.}\ \bibnamefont {Wong}},
  \bibinfo {author} {\bibfnamefont {R.-M.}\ \bibnamefont {Ma}}, \bibinfo
  {author} {\bibfnamefont {Y.}~\bibnamefont {Wang}},\ and\ \bibinfo {author}
  {\bibfnamefont {X.}~\bibnamefont {Zhang}},\ }\bibfield  {title} {\bibinfo
  {title} {Single-mode laser by parity-time symmetry breaking},\ }\href@noop {}
  {\bibfield  {journal} {\bibinfo  {journal} {Science}\ }\textbf {\bibinfo
  {volume} {346}},\ \bibinfo {pages} {972} (\bibinfo {year}
  {2014})}\BibitemShut {NoStop}%
\bibitem [{\citenamefont {Sakhdari}\ \emph {et~al.}(2017)\citenamefont
  {Sakhdari}, \citenamefont {Farhat},\ and\ \citenamefont
  {Chen}}]{sakhdari2017pt}%
  \BibitemOpen
  \bibfield  {author} {\bibinfo {author} {\bibfnamefont {M.}~\bibnamefont
  {Sakhdari}}, \bibinfo {author} {\bibfnamefont {M.}~\bibnamefont {Farhat}},\
  and\ \bibinfo {author} {\bibfnamefont {P.-Y.}\ \bibnamefont {Chen}},\
  }\bibfield  {title} {\bibinfo {title} {{PT}-symmetric metasurfaces: wave
  manipulation and sensing using singular points},\ }\href@noop {} {\bibfield
  {journal} {\bibinfo  {journal} {New Journal of Physics}\ }\textbf {\bibinfo
  {volume} {19}},\ \bibinfo {pages} {065002} (\bibinfo {year}
  {2017})}\BibitemShut {NoStop}%
\bibitem [{\citenamefont {Chen}\ \emph {et~al.}(2018)\citenamefont {Chen},
  \citenamefont {Sakhdari}, \citenamefont {Hajizadegan}, \citenamefont {Cui},
  \citenamefont {Cheng}, \citenamefont {El-Ganainy},\ and\ \citenamefont
  {Al{\`u}}}]{chen2018generalized}%
  \BibitemOpen
  \bibfield  {author} {\bibinfo {author} {\bibfnamefont {P.-Y.}\ \bibnamefont
  {Chen}}, \bibinfo {author} {\bibfnamefont {M.}~\bibnamefont {Sakhdari}},
  \bibinfo {author} {\bibfnamefont {M.}~\bibnamefont {Hajizadegan}}, \bibinfo
  {author} {\bibfnamefont {Q.}~\bibnamefont {Cui}}, \bibinfo {author}
  {\bibfnamefont {M.~M.-C.}\ \bibnamefont {Cheng}}, \bibinfo {author}
  {\bibfnamefont {R.}~\bibnamefont {El-Ganainy}},\ and\ \bibinfo {author}
  {\bibfnamefont {A.}~\bibnamefont {Al{\`u}}},\ }\bibfield  {title} {\bibinfo
  {title} {Generalized parity--time symmetry condition for enhanced sensor
  telemetry},\ }\href@noop {} {\bibfield  {journal} {\bibinfo  {journal}
  {Nature Electronics}\ }\textbf {\bibinfo {volume} {1}},\ \bibinfo {pages}
  {297} (\bibinfo {year} {2018})}\BibitemShut {NoStop}%
\bibitem [{\citenamefont {Heiss}(2004)}]{heiss2004exceptional}%
  \BibitemOpen
  \bibfield  {author} {\bibinfo {author} {\bibfnamefont {W.}~\bibnamefont
  {Heiss}},\ }\bibfield  {title} {\bibinfo {title} {Exceptional points of
  non-hermitian operators},\ }\href@noop {} {\bibfield  {journal} {\bibinfo
  {journal} {Journal of Physics A: Mathematical and General}\ }\textbf
  {\bibinfo {volume} {37}},\ \bibinfo {pages} {2455} (\bibinfo {year}
  {2004})}\BibitemShut {NoStop}%
\bibitem [{\citenamefont {Vishik}\ and\ \citenamefont
  {Lyusternik}(1960)}]{vishik1960solution}%
  \BibitemOpen
  \bibfield  {author} {\bibinfo {author} {\bibfnamefont {M.~I.}\ \bibnamefont
  {Vishik}}\ and\ \bibinfo {author} {\bibfnamefont {L.~A.}\ \bibnamefont
  {Lyusternik}},\ }\bibfield  {title} {\bibinfo {title} {The solution of some
  perturbation problems for matrices and selfadjoint or non-selfadjoint
  differential equations i},\ }\href@noop {} {\bibfield  {journal} {\bibinfo
  {journal} {Russian Mathematical Surveys}\ }\textbf {\bibinfo {volume} {15}},\
  \bibinfo {pages} {1} (\bibinfo {year} {1960})}\BibitemShut {NoStop}%
\bibitem [{\citenamefont {Lancaster}(1964)}]{lancaster1964eigenvalues}%
  \BibitemOpen
  \bibfield  {author} {\bibinfo {author} {\bibfnamefont {P.}~\bibnamefont
  {Lancaster}},\ }\bibfield  {title} {\bibinfo {title} {On eigenvalues of
  matrices dependent on a parameter},\ }\href@noop {} {\bibfield  {journal}
  {\bibinfo  {journal} {Numerische Mathematik}\ }\textbf {\bibinfo {volume}
  {6}},\ \bibinfo {pages} {377} (\bibinfo {year} {1964})}\BibitemShut {NoStop}%
\bibitem [{\citenamefont {Kato}(1966)}]{kato1966perturbation}%
  \BibitemOpen
  \bibfield  {author} {\bibinfo {author} {\bibfnamefont {T.}~\bibnamefont
  {Kato}},\ }\href@noop {} {\emph {\bibinfo {title} {Perturbation theory for
  linear operators}}}\ (\bibinfo  {publisher} {Springer-Verlag New York Inc.,
  New York},\ \bibinfo {year} {1966})\BibitemShut {NoStop}%
\bibitem [{\citenamefont {Seyranian}(1993)}]{seyranian1993sensitivity}%
  \BibitemOpen
  \bibfield  {author} {\bibinfo {author} {\bibfnamefont {A.~P.}\ \bibnamefont
  {Seyranian}},\ }\bibfield  {title} {\bibinfo {title} {Sensitivity analysis of
  multiple eigenvalues},\ }\href@noop {} {\bibfield  {journal} {\bibinfo
  {journal} {Journal of Structural Mechanics}\ }\textbf {\bibinfo {volume}
  {21}},\ \bibinfo {pages} {261} (\bibinfo {year} {1993})}\BibitemShut
  {NoStop}%
\bibitem [{\citenamefont {Bender}\ and\ \citenamefont
  {Boettcher}(1998)}]{bender1998real}%
  \BibitemOpen
  \bibfield  {author} {\bibinfo {author} {\bibfnamefont {C.~M.}\ \bibnamefont
  {Bender}}\ and\ \bibinfo {author} {\bibfnamefont {S.}~\bibnamefont
  {Boettcher}},\ }\bibfield  {title} {\bibinfo {title} {Real spectra in
  non-hermitian hamiltonians having p t symmetry},\ }\href@noop {} {\bibfield
  {journal} {\bibinfo  {journal} {Physical review letters}\ }\textbf {\bibinfo
  {volume} {80}},\ \bibinfo {pages} {5243} (\bibinfo {year}
  {1998})}\BibitemShut {NoStop}%
\bibitem [{\citenamefont {Berry}(2004)}]{berry2004physics}%
  \BibitemOpen
  \bibfield  {author} {\bibinfo {author} {\bibfnamefont {M.~V.}\ \bibnamefont
  {Berry}},\ }\bibfield  {title} {\bibinfo {title} {Physics of nonhermitian
  degeneracies},\ }\href@noop {} {\bibfield  {journal} {\bibinfo  {journal}
  {Czechoslovak journal of physics}\ }\textbf {\bibinfo {volume} {54}},\
  \bibinfo {pages} {1039} (\bibinfo {year} {2004})}\BibitemShut {NoStop}%
\bibitem [{\citenamefont {Xu}\ \emph {et~al.}(2016)\citenamefont {Xu},
  \citenamefont {Chen}, \citenamefont {Ren}, \citenamefont {Wong},\ and\
  \citenamefont {Chi}}]{xu2016self}%
  \BibitemOpen
  \bibfield  {author} {\bibinfo {author} {\bibfnamefont {L.}~\bibnamefont
  {Xu}}, \bibinfo {author} {\bibfnamefont {Q.}~\bibnamefont {Chen}}, \bibinfo
  {author} {\bibfnamefont {X.}~\bibnamefont {Ren}}, \bibinfo {author}
  {\bibfnamefont {S.-C.}\ \bibnamefont {Wong}},\ and\ \bibinfo {author}
  {\bibfnamefont {K.~T.}\ \bibnamefont {Chi}},\ }\bibfield  {title} {\bibinfo
  {title} {Self-oscillating resonant converter with contactless power transfer
  and integrated current sensing transformer},\ }\href@noop {} {\bibfield
  {journal} {\bibinfo  {journal} {IEEE Transactions on Power Electronics}\
  }\textbf {\bibinfo {volume} {32}},\ \bibinfo {pages} {4839} (\bibinfo {year}
  {2016})}\BibitemShut {NoStop}%
\bibitem [{\citenamefont {Ra’Di}\ \emph {et~al.}(2018)\citenamefont
  {Ra’Di}, \citenamefont {Chowkwale}, \citenamefont {Valagiannopoulos},
  \citenamefont {Liu}, \citenamefont {Alu}, \citenamefont {Simovski},\ and\
  \citenamefont {Tretyakov}}]{ra2018site}%
  \BibitemOpen
  \bibfield  {author} {\bibinfo {author} {\bibfnamefont {Y.}~\bibnamefont
  {Ra’Di}}, \bibinfo {author} {\bibfnamefont {B.}~\bibnamefont {Chowkwale}},
  \bibinfo {author} {\bibfnamefont {C.}~\bibnamefont {Valagiannopoulos}},
  \bibinfo {author} {\bibfnamefont {F.}~\bibnamefont {Liu}}, \bibinfo {author}
  {\bibfnamefont {A.}~\bibnamefont {Alu}}, \bibinfo {author} {\bibfnamefont
  {C.~R.}\ \bibnamefont {Simovski}},\ and\ \bibinfo {author} {\bibfnamefont
  {S.~A.}\ \bibnamefont {Tretyakov}},\ }\bibfield  {title} {\bibinfo {title}
  {On-site wireless power generation},\ }\href@noop {} {\bibfield  {journal}
  {\bibinfo  {journal} {IEEE Transactions on Antennas and Propagation}\
  }\textbf {\bibinfo {volume} {66}},\ \bibinfo {pages} {4260} (\bibinfo {year}
  {2018})}\BibitemShut {NoStop}%
\bibitem [{\citenamefont {Liu}\ \emph {et~al.}(2019)\citenamefont {Liu},
  \citenamefont {Chowkwale}, \citenamefont {Jayathurathnage},\ and\
  \citenamefont {Tretyakov}}]{liu2019pulsed}%
  \BibitemOpen
  \bibfield  {author} {\bibinfo {author} {\bibfnamefont {F.}~\bibnamefont
  {Liu}}, \bibinfo {author} {\bibfnamefont {B.}~\bibnamefont {Chowkwale}},
  \bibinfo {author} {\bibfnamefont {P.}~\bibnamefont {Jayathurathnage}},\ and\
  \bibinfo {author} {\bibfnamefont {S.}~\bibnamefont {Tretyakov}},\ }\bibfield
  {title} {\bibinfo {title} {Pulsed self-oscillating nonlinear systems for
  robust wireless power transfer},\ }\href@noop {} {\bibfield  {journal}
  {\bibinfo  {journal} {Physical Review Applied}\ }\textbf {\bibinfo {volume}
  {12}},\ \bibinfo {pages} {054040} (\bibinfo {year} {2019})}\BibitemShut
  {NoStop}%
\bibitem [{\citenamefont {Feng}\ and\ \citenamefont
  {Sit}(2020)}]{feng2020injection}%
  \BibitemOpen
  \bibfield  {author} {\bibinfo {author} {\bibfnamefont {G.}~\bibnamefont
  {Feng}}\ and\ \bibinfo {author} {\bibfnamefont {J.-J.}\ \bibnamefont {Sit}},\
  }\bibfield  {title} {\bibinfo {title} {An injection-locked wireless power
  transfer transmitter with automatic maximum efficiency tracking},\
  }\href@noop {} {\bibfield  {journal} {\bibinfo  {journal} {IEEE Transactions
  on Industrial Electronics}\ }\textbf {\bibinfo {volume} {68}},\ \bibinfo
  {pages} {5733} (\bibinfo {year} {2020})}\BibitemShut {NoStop}%
\bibitem [{\citenamefont {Wu}\ \emph {et~al.}(2022)\citenamefont {Wu},
  \citenamefont {Kang},\ and\ \citenamefont {Werner}}]{wu2022generalized}%
  \BibitemOpen
  \bibfield  {author} {\bibinfo {author} {\bibfnamefont {Y.}~\bibnamefont
  {Wu}}, \bibinfo {author} {\bibfnamefont {L.}~\bibnamefont {Kang}},\ and\
  \bibinfo {author} {\bibfnamefont {D.~H.}\ \bibnamefont {Werner}},\ }\bibfield
   {title} {\bibinfo {title} {Generalized {PT} symmetry in non-hermitian
  wireless power transfer systems},\ }\href@noop {} {\bibfield  {journal}
  {\bibinfo  {journal} {Physical review letters}\ }\textbf {\bibinfo {volume}
  {129}},\ \bibinfo {pages} {200201} (\bibinfo {year} {2022})}\BibitemShut
  {NoStop}%
\bibitem [{\citenamefont {Hao}\ \emph {et~al.}(2023)\citenamefont {Hao},
  \citenamefont {Yin}, \citenamefont {Zou}, \citenamefont {Wang}, \citenamefont
  {Huang}, \citenamefont {Ma},\ and\ \citenamefont {Dong}}]{hao2023frequency}%
  \BibitemOpen
  \bibfield  {author} {\bibinfo {author} {\bibfnamefont {X.}~\bibnamefont
  {Hao}}, \bibinfo {author} {\bibfnamefont {K.}~\bibnamefont {Yin}}, \bibinfo
  {author} {\bibfnamefont {J.}~\bibnamefont {Zou}}, \bibinfo {author}
  {\bibfnamefont {R.}~\bibnamefont {Wang}}, \bibinfo {author} {\bibfnamefont
  {Y.}~\bibnamefont {Huang}}, \bibinfo {author} {\bibfnamefont
  {X.}~\bibnamefont {Ma}},\ and\ \bibinfo {author} {\bibfnamefont
  {T.}~\bibnamefont {Dong}},\ }\bibfield  {title} {\bibinfo {title}
  {Frequency-stable robust wireless power transfer based on high-order
  pseudo-hermitian physics},\ }\href@noop {} {\bibfield  {journal} {\bibinfo
  {journal} {Physical Review Letters}\ }\textbf {\bibinfo {volume} {130}},\
  \bibinfo {pages} {077202} (\bibinfo {year} {2023})}\BibitemShut {NoStop}%
\bibitem [{\citenamefont {Guo}\ \emph {et~al.}(2024)\citenamefont {Guo},
  \citenamefont {Yang}, \citenamefont {Zhang}, \citenamefont {Wu},
  \citenamefont {Wu}, \citenamefont {Zhu}, \citenamefont {Jiang}, \citenamefont
  {Jiang}, \citenamefont {Yang}, \citenamefont {Li} \emph
  {et~al.}}]{guo2024level}%
  \BibitemOpen
  \bibfield  {author} {\bibinfo {author} {\bibfnamefont {Z.}~\bibnamefont
  {Guo}}, \bibinfo {author} {\bibfnamefont {F.}~\bibnamefont {Yang}}, \bibinfo
  {author} {\bibfnamefont {H.}~\bibnamefont {Zhang}}, \bibinfo {author}
  {\bibfnamefont {X.}~\bibnamefont {Wu}}, \bibinfo {author} {\bibfnamefont
  {Q.}~\bibnamefont {Wu}}, \bibinfo {author} {\bibfnamefont {K.}~\bibnamefont
  {Zhu}}, \bibinfo {author} {\bibfnamefont {J.}~\bibnamefont {Jiang}}, \bibinfo
  {author} {\bibfnamefont {H.}~\bibnamefont {Jiang}}, \bibinfo {author}
  {\bibfnamefont {Y.}~\bibnamefont {Yang}}, \bibinfo {author} {\bibfnamefont
  {Y.}~\bibnamefont {Li}}, \emph {et~al.},\ }\bibfield  {title} {\bibinfo
  {title} {Level pinning of anti-{PT}-symmetric circuits for efficient wireless
  power transfer},\ }\href@noop {} {\bibfield  {journal} {\bibinfo  {journal}
  {National Science Review}\ }\textbf {\bibinfo {volume} {11}},\ \bibinfo
  {pages} {nwad172} (\bibinfo {year} {2024})}\BibitemShut {NoStop}%
\bibitem [{\citenamefont {Ardila}\ \emph {et~al.}(2023)\citenamefont {Ardila},
  \citenamefont {Ram{\'\i}rez},\ and\ \citenamefont
  {Su{\'a}rez}}]{ardila2023analysis}%
  \BibitemOpen
  \bibfield  {author} {\bibinfo {author} {\bibfnamefont {V.}~\bibnamefont
  {Ardila}}, \bibinfo {author} {\bibfnamefont {F.}~\bibnamefont
  {Ram{\'\i}rez}},\ and\ \bibinfo {author} {\bibfnamefont {A.}~\bibnamefont
  {Su{\'a}rez}},\ }\bibfield  {title} {\bibinfo {title} {Analysis of an
  oscillatory system with three coupled coils for wireless power transfer},\
  }\href@noop {} {\bibfield  {journal} {\bibinfo  {journal} {IEEE Transactions
  on Microwave Theory and Techniques}\ }\textbf {\bibinfo {volume} {72}},\
  \bibinfo {pages} {3387} (\bibinfo {year} {2023})}\BibitemShut {NoStop}%
\bibitem [{\citenamefont {Haus}(1984)}]{haus1984optoelectronics}%
  \BibitemOpen
  \bibfield  {author} {\bibinfo {author} {\bibfnamefont {H.~A.}\ \bibnamefont
  {Haus}},\ }\href@noop {} {\emph {\bibinfo {title} {Waves and fields in
  optoelectronics}}}\ (\bibinfo  {publisher} {Prentice-Hall, Englewood Cliffs,
  NJ},\ \bibinfo {year} {1984})\BibitemShut {NoStop}%
\bibitem [{\citenamefont {Zhou}\ \emph {et~al.}(2018)\citenamefont {Zhou},
  \citenamefont {Zhang}, \citenamefont {Xiao}, \citenamefont {Qiu},\ and\
  \citenamefont {Chen}}]{zhou2018nonlinear}%
  \BibitemOpen
  \bibfield  {author} {\bibinfo {author} {\bibfnamefont {J.}~\bibnamefont
  {Zhou}}, \bibinfo {author} {\bibfnamefont {B.}~\bibnamefont {Zhang}},
  \bibinfo {author} {\bibfnamefont {W.}~\bibnamefont {Xiao}}, \bibinfo {author}
  {\bibfnamefont {D.}~\bibnamefont {Qiu}},\ and\ \bibinfo {author}
  {\bibfnamefont {Y.}~\bibnamefont {Chen}},\ }\bibfield  {title} {\bibinfo
  {title} {Nonlinear parity-time-symmetric model for constant efficiency
  wireless power transfer: Application to a drone-in-flight wireless charging
  platform},\ }\href@noop {} {\bibfield  {journal} {\bibinfo  {journal} {IEEE
  Transactions on Industrial Electronics}\ }\textbf {\bibinfo {volume} {66}},\
  \bibinfo {pages} {4097} (\bibinfo {year} {2018})}\BibitemShut {NoStop}%
\bibitem [{\citenamefont {Abdelshafy}\ \emph {et~al.}(2019)\citenamefont
  {Abdelshafy}, \citenamefont {Othman}, \citenamefont {Oshmarin}, \citenamefont
  {Almutawa},\ and\ \citenamefont {Capolino}}]{abdelshafy2019exceptional}%
  \BibitemOpen
  \bibfield  {author} {\bibinfo {author} {\bibfnamefont {A.~F.}\ \bibnamefont
  {Abdelshafy}}, \bibinfo {author} {\bibfnamefont {M.~A.}\ \bibnamefont
  {Othman}}, \bibinfo {author} {\bibfnamefont {D.}~\bibnamefont {Oshmarin}},
  \bibinfo {author} {\bibfnamefont {A.~T.}\ \bibnamefont {Almutawa}},\ and\
  \bibinfo {author} {\bibfnamefont {F.}~\bibnamefont {Capolino}},\ }\bibfield
  {title} {\bibinfo {title} {Exceptional points of degeneracy in periodic
  coupled waveguides and the interplay of gain and radiation loss: Theoretical
  and experimental demonstration},\ }\href@noop {} {\bibfield  {journal}
  {\bibinfo  {journal} {IEEE Transactions on Antennas and Propagation}\
  }\textbf {\bibinfo {volume} {67}},\ \bibinfo {pages} {6909} (\bibinfo {year}
  {2019})}\BibitemShut {NoStop}%
\bibitem [{\citenamefont {Nikzamir}\ and\ \citenamefont
  {Capolino}(2022)}]{nikzamir2022highly}%
  \BibitemOpen
  \bibfield  {author} {\bibinfo {author} {\bibfnamefont {A.}~\bibnamefont
  {Nikzamir}}\ and\ \bibinfo {author} {\bibfnamefont {F.}~\bibnamefont
  {Capolino}},\ }\bibfield  {title} {\bibinfo {title} {Highly sensitive coupled
  oscillator based on an exceptional point of degeneracy and nonlinearity},\
  }\href@noop {} {\bibfield  {journal} {\bibinfo  {journal} {Physical Review
  Applied}\ }\textbf {\bibinfo {volume} {18}},\ \bibinfo {pages} {054059}
  (\bibinfo {year} {2022})}\BibitemShut {NoStop}%
\bibitem [{\citenamefont {Kurs}\ \emph {et~al.}(2010)\citenamefont {Kurs},
  \citenamefont {Moffatt},\ and\ \citenamefont
  {Solja{\v{c}}i{\'c}}}]{kurs2010simultaneous}%
  \BibitemOpen
  \bibfield  {author} {\bibinfo {author} {\bibfnamefont {A.}~\bibnamefont
  {Kurs}}, \bibinfo {author} {\bibfnamefont {R.}~\bibnamefont {Moffatt}},\ and\
  \bibinfo {author} {\bibfnamefont {M.}~\bibnamefont {Solja{\v{c}}i{\'c}}},\
  }\bibfield  {title} {\bibinfo {title} {Simultaneous mid-range power transfer
  to multiple devices},\ }\href@noop {} {\bibfield  {journal} {\bibinfo
  {journal} {Applied Physics Letters}\ }\textbf {\bibinfo {volume} {96}}
  (\bibinfo {year} {2010})}\BibitemShut {NoStop}%
\bibitem [{\citenamefont {Oshmarin}\ \emph {et~al.}(2021)\citenamefont
  {Oshmarin}, \citenamefont {Abdelshafy}, \citenamefont {Nikzamir},
  \citenamefont {Green},\ and\ \citenamefont
  {Capolino}}]{oshmarin2021experimental}%
  \BibitemOpen
  \bibfield  {author} {\bibinfo {author} {\bibfnamefont {D.}~\bibnamefont
  {Oshmarin}}, \bibinfo {author} {\bibfnamefont {A.~F.}\ \bibnamefont
  {Abdelshafy}}, \bibinfo {author} {\bibfnamefont {A.}~\bibnamefont
  {Nikzamir}}, \bibinfo {author} {\bibfnamefont {M.~M.}\ \bibnamefont
  {Green}},\ and\ \bibinfo {author} {\bibfnamefont {F.}~\bibnamefont
  {Capolino}},\ }\bibfield  {title} {\bibinfo {title} {Experimental
  demonstration of a new oscillator concept based on degenerate band edge in
  microstrip circuit},\ }\href@noop {} {\bibfield  {journal} {\bibinfo
  {journal} {arXiv preprint arXiv:2109.07002}\ } (\bibinfo {year}
  {2021})}\BibitemShut {NoStop}%
\bibitem [{\citenamefont {Zhou}\ and\ \citenamefont
  {Chong}(2016)}]{zhou2016pt}%
  \BibitemOpen
  \bibfield  {author} {\bibinfo {author} {\bibfnamefont {X.}~\bibnamefont
  {Zhou}}\ and\ \bibinfo {author} {\bibfnamefont {Y.}~\bibnamefont {Chong}},\
  }\bibfield  {title} {\bibinfo {title} {{PT} symmetry breaking and nonlinear
  optical isolation in coupled microcavities},\ }\href@noop {} {\bibfield
  {journal} {\bibinfo  {journal} {Optics express}\ }\textbf {\bibinfo {volume}
  {24}},\ \bibinfo {pages} {6916} (\bibinfo {year} {2016})}\BibitemShut
  {NoStop}%
\end{thebibliography}


\end{document}